\documentclass[journal]{IEEEtran}

\usepackage[colorlinks,linkcolor=red,hyperindex,bookmarks=false]{hyperref}

\usepackage{ifpdf}
\usepackage{cite}
\usepackage{algorithm}  
\usepackage{algpseudocode}  
\usepackage{amsmath}  

\usepackage{indentfirst}
\usepackage{graphicx}
\usepackage[justification=centering]{caption}
\usepackage{color}
\usepackage[american]{babel}
\usepackage{microtype}
\usepackage{lipsum}
\usepackage[caption=false, font=footnotesize]{subfig}
\usepackage{makecell}
\usepackage{setspace}
\usepackage{bm}
\usepackage{multirow}
\usepackage{soul}
\soulregister\cite7
\soulregister\ref7

\ifCLASSINFOpdf
\else
 \DeclareGraphicsExtensions{.png}
\fi

\usepackage{array}
\usepackage{stfloats}
\usepackage{url}

\begin{document}

\title{LCSCNet: Linear Compressing Based Skip-Connecting Network\\ for Image Super-Resolution}

\author{Wenming~Yang,
        Xuechen~Zhang,
        Yapeng~Tian,
        Wei~Wang,
        Jing-Hao~Xue,
        and~Qingmin~Liao
        \thanks{W. Yang, X. Zhang, W. Wang and Q. Liao are with the Shenzhen Key Lab of Information Science and Technology, Shenzhen Engineering Lab of IS\&DRM, Department of Electronic Engineering, Graduate School at Shenzhen, Tsinghua University, Shenzhen 518055, China (E-mail: \{yang.wenming@sz, xc-zhang16@mails, wangwei17@mails, liaoqm@\}.tsinghua.edu.cn).}
        \thanks {Y.~Tian is with the Department of Computer Science, University of Rochester, USA (E-mail: ytian21@ur.rochester.edu).}
        \thanks {J.-H.~Xue is with the Department of Statistical Science, University College London, UK (E-mail: jinghao.xue@ucl.ac.uk).}
        \thanks{This work was partly supported by the National Natural Science Foundation of China (No.61471216 and No.61771276), the National Key Research and Development Program of China (No.2016YFB0101001) and the Special Foundation for the Development of Strategic Emerging Industries of Shenzhen (No.JCYJ20170307153940960 and No.JCYJ20170817161845824).}       
}

\maketitle

\begin{abstract}
In this paper, we develop a concise but efficient network architecture called linear compressing based skip-connecting network (LCSCNet) for image super-resolution. Compared with two representative network architectures with skip connections, ResNet and DenseNet, a linear compressing layer is designed in LCSCNet for skip connection, which connects former feature maps and distinguishes them from newly-explored feature maps.
In this way, the proposed LCSCNet enjoys the merits of the distinguish feature treatment of DenseNet and the parameter-economic form of ResNet. Moreover, to better exploit hierarchical information from both low and high levels of various receptive fields in deep models, inspired by gate units in LSTM, we also propose an adaptive element-wise fusion strategy with multi-supervised training. Experimental results in comparison with state-of-the-art algorithms validate the effectiveness of LCSCNet.
\end{abstract}


\IEEEpeerreviewmaketitle

\section{Introduction}\label{s:s_1}

\IEEEPARstart{S}{ingle} image super-resolution (SISR) is a classical but challenging ill-posed inverse problem in low-level computer vision, aiming at restoring a high-resolution (HR) image from a single low-resolution (LR) input image. It is widely used in various areas such as medical imaging, satellite imaging and security imaging~\cite{yang2014single,park2003super}. 

Early methods for SISR are mainly interpolation-based, including Bicubic interpolation~\cite{keys1981cubic} and Lanczos resampling~\cite{duchon1979lanczos}. Then more powerful reconstruction-based methods often adopt sophisticated prior knowledge to restrict the possible solution space, with the advantage of generating flexible and sharp details~\cite{dai2009softcuts,sun2008image,yan2015single,marquina2008image}. Learning-based methods are now mainstream algorithms for SISR, utilizing substantial data to learn statistical relationships between LR and HR pairs. Markov Random Field (MRF)~\cite{freeman2002example} was firstly adopted by 
Freeman \emph{et~al.} to exploit the abundant real-world images to synthesize visually pleasing image textures. Neighbor embedding methods~\cite{chang2004super} proposed by Chang \emph{et~al.} took advantage of similar local geometry between LR and HR to restore HR image patches. Inspired by the sparse signal recovery theory, researchers applied sparse coding methods~\cite{yang2010image,zeyde2010single,timofte2013anchored,timofte2014a+,yang2016consistent} to SR. Random forest~\cite{schulter2015fast} has also been used to improve the reconstruction performance. 

Recently, remarkable performance has been achieved for SR by deep models, especially deep network architectures, which are elaborated for high-level tasks in computer vision. Notably, residual network (ResNet) and densely connected network (DenseNet) are two widely-used architectures, which use skip connections to alleviate gradient problems and degradation phenomena in training. Chen \emph{et~al.}~\cite{chen2017dual} analyzed ResNet and DenseNet in the HORNN framework~\cite{soltani2016higher} and concluded that ResNet enables feature re-usage while DenseNet enables feature exploration, both important to learn powerful representations.

Through extensive experiments, \cite{veit2016residual} and~\cite{huang2016deep} implied that ResNet shows an ensemble-like behavior within its structure. Yang \emph{et~al.}~\cite{yang2017deep} showed that ResNet applied in SR would lead to output with a layer-by-layer progressive effect, and Huang \emph{et~al.}~\cite{huang2017densely} argued that this might restrict ResNet from reaching more feasible solutions. Although DenseNet explores as many new features as possible by directly utilizing any former features, its excessive skip connections among intermediate layers increase the number of parameters and burden the hardware during training.

In this paper, we propose Linear Compressing Based Skip-Connecting Network (LCSCNet), as a framework for SR, which takes advantages of ResNet's parameter-economic feature re-usage and DenseNet's distinguishing feature exploration, as well as mitigating difficulties of restricted structures of ResNet and parameter burden of DenseNet.

As the network depth grows, the features produced by different intermediate layers would be hierarchical with different receptive fields. Among deep SR models, DRCN~\cite{kim2016deeply} and MemNet~\cite{Tai-MemNet-2017} used these intermediate features with multi-supervised methods, in which each feature corresponded to a raw SR output, and then fused these intermediate SR outputs by a list of trained scalars. Such a fusion strategy has two flaws: 1) once the weight scalars are determined in training, it will not change with different inputs; 2) using a single scalar to weight SR output fails to take pixel-wise differences into consideration, i.e., it would be better to weight different parts distinguishingly in an adaptive way. 
To overcome these shortcomings, inspired by the gate units in LSTM~\cite{hochreiter1997long}, we develop an adaptive element-wise fusion strategy in a progressive constructive way to maintain the element-wise convex weighted pattern, aiming at making better use of hierarchical information with different receptive fields.

In the end, we composite the Basic LCSCNet architecture with the adaptive element-wise fusion strategy gracefully for SR. Analysis and experiments in the following sections will illustrate the rationality of the proposed methods. 

The main contributions of this work are three-fold:

1) We propose an accurate and efficient Linear Compressing Based Skip-Connecting Network (LCSCNet) architecture, which inherits the advantage of DenseNet in treating features of different levels distinguishingly while reducing its parameter size by exploiting the parameter-economic strength of ResNet. Moreover, we develop an Enhanced LCSCNet (E-LCSCNet) to further alleviate difficulties of training large-scale networks.

2) Differently from the traditional stationary fusion strategy, we take the input differences as well as the element-wise variation into consideration and propose an adaptive element-wise fusion strategy to further utilize hierarchical information.

3) When compared with the state-of-the-art models trained on the widely-used 291 dataset and those light networks trained on the DIV2K dataset, our proposed framework achieves the state-of-the-art performance. When compared with large models trained on DIV2K, our E-LCSCNet is among the state-of-the-art with apparent parametric efficiency.

The rest of the paper is organized as follows. Section~\ref{s:s_2} reviews recent related work. Section~\ref{s:s_3} presents a detailed description of the proposed architecture, mainly on the configuration of Basic LCSCNet and the adaptive element-wise fusion algorithm. Section~\ref{s:s_4} illustrates several intriguing properties of LCSCNet, which could explain the rationality of LCSCNet. Section~\ref{s:s_5} conducts ablation studies to further probe into the proposed framework. Section~\ref{s:s_6} presents experimental results in comparison with other relevant methods. 
Section~\ref{s:s_8} concludes the paper and envisages some future work.

\begin{figure*}[htbp]
\centering
\subfloat[Overall architecture of Basic LCSCNet (E-LCSCNet)]{
\includegraphics[scale=0.5]{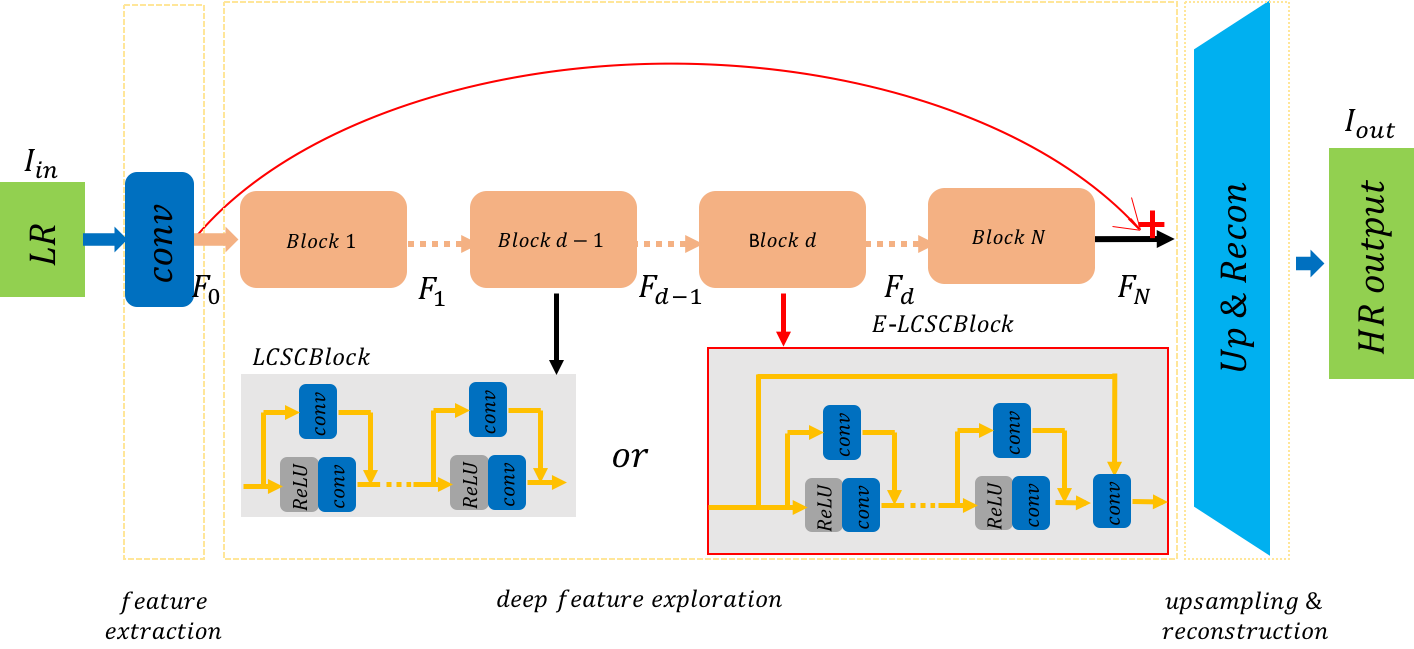}} \\
\subfloat[Overall architecture of LCSCNet (E-LCSCNet)]{
\includegraphics[scale=0.5]{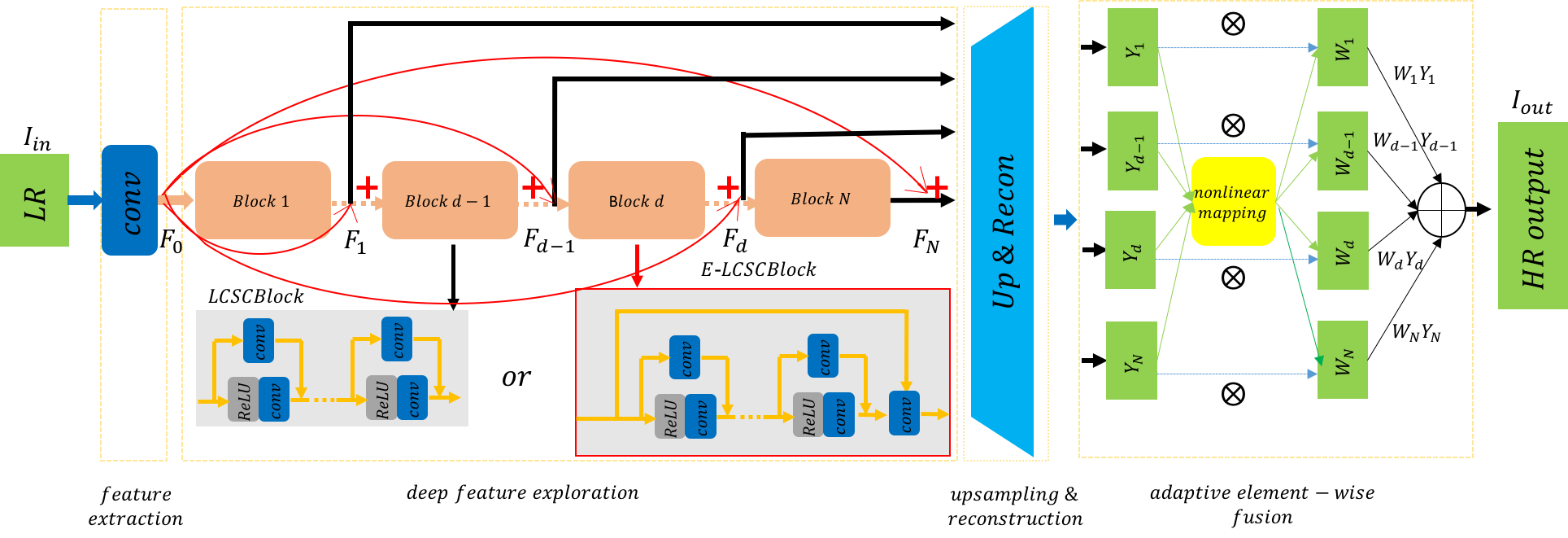}}
\caption{\small The overall architectures of (a) Basic LCSCNet (E-LCSCNet) and (b) LCSCNet (E-LCSCNet). In (b), $\otimes$ means element-wise multiplication; $\{Y_{1},\dots,Y_{N}\}$ are the intermediate HR outputs reconstructed from $\{F_{1},\dots,F_{N}\}$. When E-LSCSNet is employed, red lined parts are activated. For fair comparison, the upsampling and reconstruction part of (Basic) LCSCNet varies with the training dataset: For models trained on the 291 dataset, this part is the traditional deconv layer consisting of ``nearst-neighborhood upsampling + conv-ReLU + conv-ReLU + conv"; for models trained on the DIV2K dataset, we use ESPCN~\cite{shi2016real} instead. To be specific, we only use ESPCN as U\&RNet in Section~\ref{s:s_5::E} and the E-LCSCNet in Table~\ref{chart:big}.}
\label{fig:F3}
\end{figure*}

\section{Related Work}\label{s:s_2}

Because our proposed methods include the Basic LCSCNet architecture and the adaptive element-wise fusion strategy, in this section we review related work mainly from the aspects of basic SISR reconstruction and sub-output fusion. 

\subsection{Basic SISR Reconstruction} 

Dong \emph{et~al.} pioneeringly proposed a three-layer super-resolution convolutional neural networks (SRCNN) \cite{dong2014learning}, predicting the end-to-end nonlinear mapping between LR and HR spaces. This first trial significantly outperformed other algorithms at that time. To combine the benefits of the natural sparsity of images and deep neural network architectures, Wang \emph{et~al.} proposed the Cascaded Sparse Coding Network (CSCN)~\cite{wang2015deep}, which had a higher visual quality than previous work. After SRCNN, Dong \emph{et~al.} further proposed FSRCNN~\cite{dong2016accelerating} improving SRCNN mainly by leveraging deconvolution layers, which reduced computation significantly by increasing the resolution only at the end of network. In the meantime, the Efficient Sub-Pixel Convolution Neural Network (ESPCN)~\cite{shi2016real} was proposed by Shi \emph{et~al.}, replacing the traditional deconvolution layer by an efficient sub-pixel convolution layer and further reducing computation. 

Inspired by the success that very deep neural networks with sophisticated architectures and training strategies achieved in some high-level tasks in computer vision~\cite{simonyan2014very}, Kim \emph{et~al.} employed the VGG architecture and high learning rate with gradient clipping to stack a very deep (20 layers) convolutional neural network (VDSR)~\cite{kim2016accurate} and gained a remarkable improvement. Mao \emph{et~al.} proposed a deep fully convolutional auto-encoder network with symmetric skip connections~\cite{mao2016image}. To handle the issue of large numbers of parameters brought by very deep architectures, Kim \emph{et~al.} proposed the Deeply-Recursive Convolutional Network (DRCN)~\cite{kim2016deeply}, which was also 20-layer but with 16 recursions among its intermediate layers. To further exploit the advantages from deepening neural networks, motivated by the success of~\cite{he2016deep}, Tai \emph{et~al.} proposed the Deep Recursive Residual Network (DRRN)~\cite{tai2017image}, a 54-layer convolutional neural network for SR, in which they utilized the residual network architecture (ResNet)~\cite{he2016identity} in both global and local manners. Inspired by the Dense Connected Network (DenseNet) \cite{huang2017densely} proposed by Huang \emph{et~al.}, Tong \emph{et~al.} introduced dense skip connections to their deep architecture~\cite{tong2017image}. Based on the correlations among the HR outputs with different scale factors and a heuristic methodology, Lai \emph{et~al.} proposed the Laplacian Pyramid Super-Resolution Network (LapSRN)~\cite{LapSRN} to progressively reconstruct the sub-band residuals of higher-resolution images, which was especially effective for large scale factors. Motivated by explicitly mining persistent memory through an adaptive learning process and further mitigating the difficulties of training deeper networks, Tai \emph{et~al.} proposed an 80-layer network for image restoration, named as Persistent Memory Network (MemNet)~\cite{Tai-MemNet-2017}. 

Very recently, to further explore the power of example-based SISR with abundant training data, a new dataset DIV2K~\cite{Agustsson_2017_CVPR_Workshops} consisting of 800 2K resolution images was established. Based on this powerful dataset, many new architectures were proposed for performance improvement. Among them, by removing Batch-Normalization (BN)~\cite{ioffe2015batch} and applying residual scaling, Lim \emph{et~al.} proposed the Enhanced Deep Residual Network (EDSR)~\cite{lim2017enhanced}, which significantly improved performance. Then the Deep Back-Projection Network (DBPN)~\cite{haris2018deep} was proposed by Haris \emph{et~al.} to combine the merits of deep neural networks with the back-projection procedure, proven to be very effective for large scale factors. By making full use of local and global information from deep architectures, the Residual Dense Network (RDN)~\cite{zhang2018residual} proposed by Zhang \emph{et~al.} exhibits comparable performance to EDSR, with fewer parameters.   

\subsection{Sub-output Fusion} 

Features from different depths with different receptive fields specialize in different patterns in SISR. From the perspective of ensemble learning, a better result can be acquired by adaptively fusing the outputs from different-level features. Based on this concept, several fusion strategies were proposed. Among them, two representative weighted-summation methods were the vectorized weighted fusion strategy~\cite{kim2016deeply,Tai-MemNet-2017} and MSCN~\cite{liu2016learning}. In the vectorized weighted fusion, a trainable positive vector whose $\ell_{1}$ norm is 1 is applied, and each element in this vector controls how much of the current sub-output contributes to the final one. To regularize each sub-output and stabilize training, multi-supervised training is adopted. In MSCN, an extra CNN module takes LR as input and outputs several tensors with the same shape as the HR. These tensors can be viewed as adaptive element-wise weights for raw HR outputs. Then the weight module and the basic SISR module are trained jointly by optimizing the fused results in an end-to-end manner. 

Both of the two fusing strategies above have shortcomings. The vectorized approach does not take the diversity of input and pixel-wise differences into consideration, while in MSCN the summation of coefficients at each pixel is not normalized, which is incongruous. Therefore, in this paper we aim to propose a normalized adaptive element-wise fusion strategy to overcome the shortcomings of the two previous fusion methods. 

The above-mentioned deep methods mainly minimized the mean squared error (MSE), which tended to be blurry, over-smoothing and perceptually unsatisfying, especially in the case of large scale factors. Recently, some inspiring deep learning-based works concentrated on the exploration of more effective loss functions for SR. In~\cite{bruna2015super} and~\cite{johnson2016perceptual}, the perceptual loss using high-level feature maps of VGG made HR outputs more visually pleasing; \cite{sonderby2016amortised} introduced amortized MAP inference to the loss function to get more plausible results; \cite{ledig2016photo} and~\cite{sajjadi2016enhancenet} used the adversarial loss to produce photo-realistic HR outputs. Although these methods produced high-quality images with rich texture details, the details in their outputs may be quite different from original images. As we mainly aim to develop efficient deep models with fewer pixel-wise errors, our work does not belong to this group. Readers can refer to~\cite{yang2018deep} for an elaborated survey on deep learning based SISR.

\section{Linear Compressing Based Skip-Connecting Network (LCSCNet and E-LCSCNet)}\label{s:s_3}

Our work has two main technical contributions: an (enhanced) linear compressing based skip-connecting structure for developing extremely deep efficient neural networks, and an adaptive fusion strategy for further utilizing intermediate features. In order to better clarify the contribution and function for each of them, here we briefly specify four architectures used in later discussions and ablation studies: 

\emph{\textbf{Basic LCSCNet:} as shown in Fig.\ref{fig:F3}(a) (without red-line parts), it firstly extracts features and then sends them to a series of LCSCBlocks, and the final results are obtained from the upsampling and reconstruction part;}

\emph{\textbf{Basic E-LCSCNet:} quite similar to Basic LCSCNet except for the replacement of LCSCBlock by E-LCSCBlock (Fig.\ref{fig:F3}(a)) and the extra additive skip connections with initial features;}

\emph{\textbf{LCSCNet} and \textbf{E-LCSCNet:} applying the proposed adaptive fusion strategy to Basic LCSCNet and Basic E-LCSCNet respectively, as shown in Fig.\ref{fig:F3}(b).} 

Because the structure of the above two basic networks are quite simple and we mainly use LCSCNet (E-LCSCNet) to compare with other state-of-the-art works, we will focus on the detailed descriptions on LCSCNet (E-LCSCNet).

As shown in Fig.\ref{fig:F3}(b), our LCSCNet and E-LCSCNet both mainly consist of four parts: 1) a preliminary feature extraction net (PFENet), 2) linear compressing based skip-connecting blocks (LCSCBlocks) or enhanced linear compressing based skip-connecting blocks (E-LCSCBlocks) for deep feature exploration, 3) a upsampling and reconstruction net (U\&RNet), and 4) an adaptive element-wise fusion of all intermediate outputs. Many previous works~\cite{kim2016accurate, tai2017image, Tai-MemNet-2017, LapSRN} learned the residue between HR and its bicubic interpolation and argued that this helps stabilize training and improves performance. When we compare LCSCNet with these works, as shown in Fig.\ref{fig:F3}, the input $I_{in}$ is LR, and the output $I_{out}$ is the residue. Meanwhile, many recent works~\cite{lim2017enhanced, zhang2018residual} just learned the mapping between LR and HR. When we compare E-LCSCNet with these methods, we also follow this routine for fairness. 

Our PFENet uses a single $3\times3$ convolution layer to conduct preliminary feature extraction:
\begin{equation}
F_{0} = f_{PFE}(I_{in}),
\label{con:over_1}
\end{equation}
where $F_{0}$ denotes extracted features from the LR input.

\begin{figure}[htbp]
\centering
\includegraphics[scale=0.38]{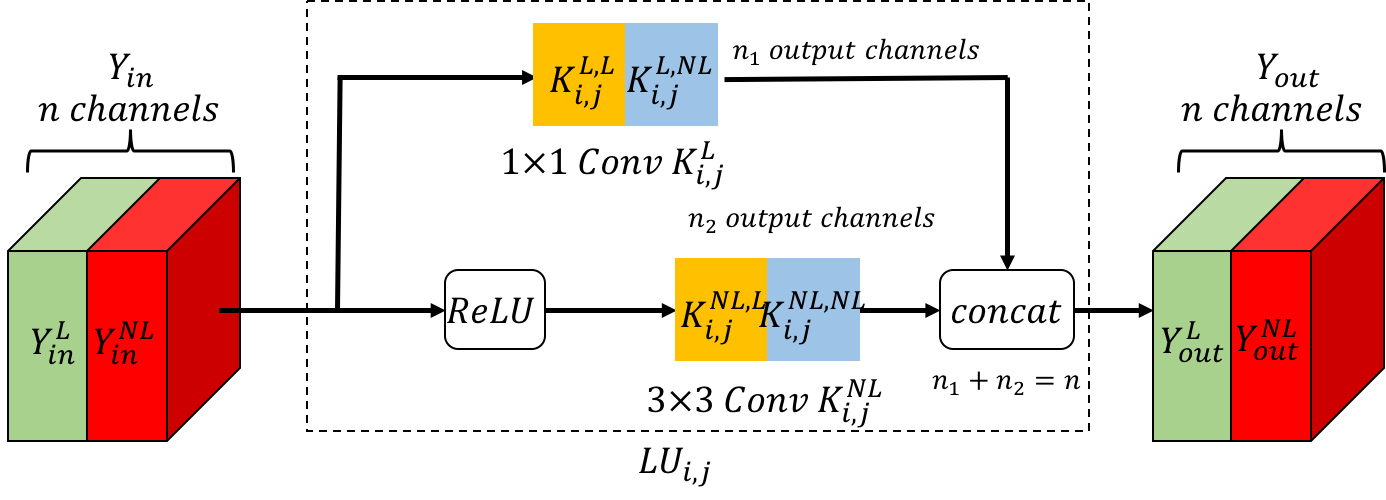}
\caption{\small The configuration of LCSCUnit.}
\label{fig:F4}
\end{figure}

\subsection{Configurations of LCSCUnit, LCSCBlock and E-LCSCBlock}

\subsubsection{LCSCUnit and LCSCBlock}

The features extracted by the PFENet are then transmitted to the second part of overall network, which uses LCSCBlocks to explore complicated features progressively. An LCSCBlock comprises a fixed number of linear compressing based skip-connecting units (LCSCUnit) with the same configuration. The basic configuration of the LCSCUnit is depicted in Fig.\ref{fig:F4}, where $LU_{i,j}$ denotes the $j$-th unit in the $i$-th LCSCBlock, $Y_{in}$ denotes the input feature maps of this unit and $Y_{out}$ denotes the output feature maps, both maps with $n$ channels. In Fig.\ref{fig:F4}, the upper convolution operator named as linear compressing (LC) layer is of size $1\times1$ with $n_{1}$ output channels. We denote the LC layer in $LU_{i,j}$ as $K^{L}_{i,j}$. Motivated by~\cite{he2016identity}, the nonlinear operator in the lower part of Fig.\ref{fig:F4} consists of two parts: ReLU and the convolution operator denoted as $K^{NL}_{i,j}$ of size $3\times3$ with $n_{2}$ output channels. Here the superscripts $^{L}$ and $^{NL}$ denote the convolution kernels for linear and nonlinear transformations, respectively. Then the output of $K^{L}_{i,j}$ and $K^{NL}_{i,j}$ are concatenated to form a $n$-channel output feature maps. For simplicity, bias is omitted and convolution is replaced by matrix multiplication\footnote{\emph{e.g.} the convolution operation X$\ast$Y is rewritten as XY for simplicity.}, then the whole process of LCSCUnit can be formulated as 
\begin{equation}
Y_{out} = concat \big (K^{L}_{i,j}Y_{in}, \ K^{NL}_{i,j}ReLU(Y_{in}) \big ).
\label{eq:lcscunit}
\end{equation}

Furthermore, features and convolution kernels in LCSCUnits can be separated by their properties. As for features, $Y_{out}$ can be divided into $n_1$-channel $Y_{out}^{L}$ and $n_2$-channel $Y_{out}^{NL}$, where superscripts $^{L}$ and $^{NL}$ in features denote features produced by linear and nonlinear operations, respectively. For convolution kernels, $K_{i,j}^{L}$ can be divided according to the output channel into $K_{i,j}^{L,L}$ and $K_{i,j}^{L,NL}$, where superscript ${}^{L,L}$ means the part of the linear-transforming kernel $K_{i,j}^{L}$ operating on $Y_{in}^{L}$ and ${}^{L,NL}$ means the part operating on $Y_{in}^{NL}$. 

Notably, although the LC layer with $1\times 1$ convolution resembles the bottleneck layer that is widely used to reduce dimensions of feature maps~\cite{lin2013network,huang2017densely}, the main difference between them is that the bottleneck layer is placed before the nonlinear operator in a cascading manner, while the LC layer parallels the nonlinear operator.

Skip connections in a neural network structure create short paths from early layers to latter layers, which are considered as an effective way to ease the difficulties in training deep neural networks. In all LCSCNet, we implement skip connections mainly by the LC layer in each basic unit. In LCSCUnit, there is a parameter which controls the proportion of the number of linear output channel $n_{1}$ and the nonlinear output channel $n_{2}$. This parameter, which can affect the performance of the network, is defined as
\begin{equation}
\rho=\frac{n_{2}}{n_{1} + n_{2}}.
\end{equation}
We find that a fixed $\rho$ for each LCSCUnit throughout the network can already offer a quite good performance. Alternatively, we can set up LCSCUnits with different $\rho$, and the LCSCUnits with the same $\rho$ are connected consecutively and can be divided into different LCSCBlocks. For simplicity, we let each LCSCBlock contain the same number of LCSCUnit.

Suppose there are $N$ LCSCBlocks stacked to explore deep features, and $M$ LCSCUnits in an LCSCBlock. Let $LB_{d}^{\rho_{d}}$ denote the $d$-th LCSCBlock with specific $\rho_{d}$, $F_{d-1}$ denote its input features and $F_{d}$ its output features. The mapping of LCSCUnits in this block are denoted by $\{LU_{d,1}(\cdot), LU_{d,2}(\cdot), \dots, LU_{d,M}(\cdot)\}$, then the whole process of this block can be formulated as 
\begin{equation}
F_{d}\!=\!LB_{d}^{\rho_{d}}(F_{d\!-\!1})\!=\!LU_{d,M}(LU_{d,M\!-\!1}(\cdots(LU_{d,1}(F_{d\!-\!1}))\cdots)),
\label{con:over_2}
\end{equation}
and it follows that
\begin{equation}
F_{N} = LB^{\rho_{N}}_{N}(LB^{\rho_{N-1}}_{N-1}(\cdots(LB^{\rho_{1}}_{1}(F_{0}))\cdots)).
\label{con:dfe_1}
\end{equation}

Furthermore, we investigate how the ordinal position of blocks with different $\rho$ effects the final performance. Detailed discussions and relative comparative experiments will be demonstrated in Section~\ref{s:s_5::C}.

\subsubsection{E-LCSCBlock}

As mentioned in~\cite{lim2017enhanced}, the simplest way to enhance performance via increasing the number of parameters is to increase the width of deep architectures. However, a deep wide network is extremely hard to train. 
Inspired by the long-term memory connection in~\cite{Tai-MemNet-2017}, we find that if we further concatenate the input and the output of LCSCBlock and then use a $1 \times 1$ bottleneck layer to maintain the compactness of the output channel, it will alleviate the difficulty of training a large LCSCNet. We denote the LCSCBlock with such a long-term memory connection as E-LCSCBlock. Compared with (\ref{con:over_2}), the mapping of E-LCSCBlock can be written as 
\begin{equation}
\begin{split}
ELB_{d}^{\rho_{d}}(F_{d-1})=bottle(concat(F_{d-1}, LB_{d}^{\rho_{d}}(F_{d-1}))),
\end{split}
\label{con:new_1}
\end{equation}
where $ELB_{d}^{\rho_{d}}$ denotes the $d$-th E-LCSCBlock with specific $\rho_{d}$, and $bottle(\cdot)$ denotes the $1 \times 1$ bottleneck layer. Moreover, we find that deep models of moderate scales using E-LCSCBlocks also perform slightly better than the ones using LCSCBlocks. Further discussions and ablation studies on E-LCSCBlock will be presented in Section~\ref{s:s_5::E}.

\subsection{Upsampling and Reconstruction Net (U\&RNet)}

Sajjadi \emph{et~al.}~\cite{sajjadi2016enhancenet} reported that adding convolution layers after the nearest-neighbor upsamping layer can help alleviate artifacts in SR. We follow this way in our models trained on the 291 dataset using the nearest-neighbor upsampling layer followed by three $3 \times 3$ convolution kernels (except the last one) with ReLU. When we develop models aiming to compare with models trained on DIV2K, we use ESPCN as the U\&RNet, as EDSR and RDN did, for fair comparison.

In LCSCNet, the deep features $\{F_{1},F_{2},\dots,F_{N}\}$, explored hierarchically in its second part by LCSCBlocks, are then sent to U\&RNet $UR(\cdot)$, which maps feature $F_{d}$ to output $Y_{d}$:
\begin{equation}
\begin{split}
Y_{d}=UR(F_{d}),~1\leq d \leq N.
\end{split}
\label{con:over_3}
\end{equation}

In E-LCSCNet, like EDSR and RDN, even without directly learning the residue between the HR and its bicubic version, the global residual learning is implemented by adding initial features $F_{0}$ to $F_{d}$ before upsampling. That is, the U\&RNet $EUR(\cdot)$ in E-LCSCNet has input and output as
\begin{equation}
    \begin{split}
        Y_{d}=EUR(F_{d}+F_{0}),~1\leq d \leq N.
    \end{split}
    \label{con:eur}
\end{equation}

\subsection{Adaptive Element-wise Fusion Strategy}

\begin{figure*}[htbp]
\centering
\includegraphics[scale=0.45]{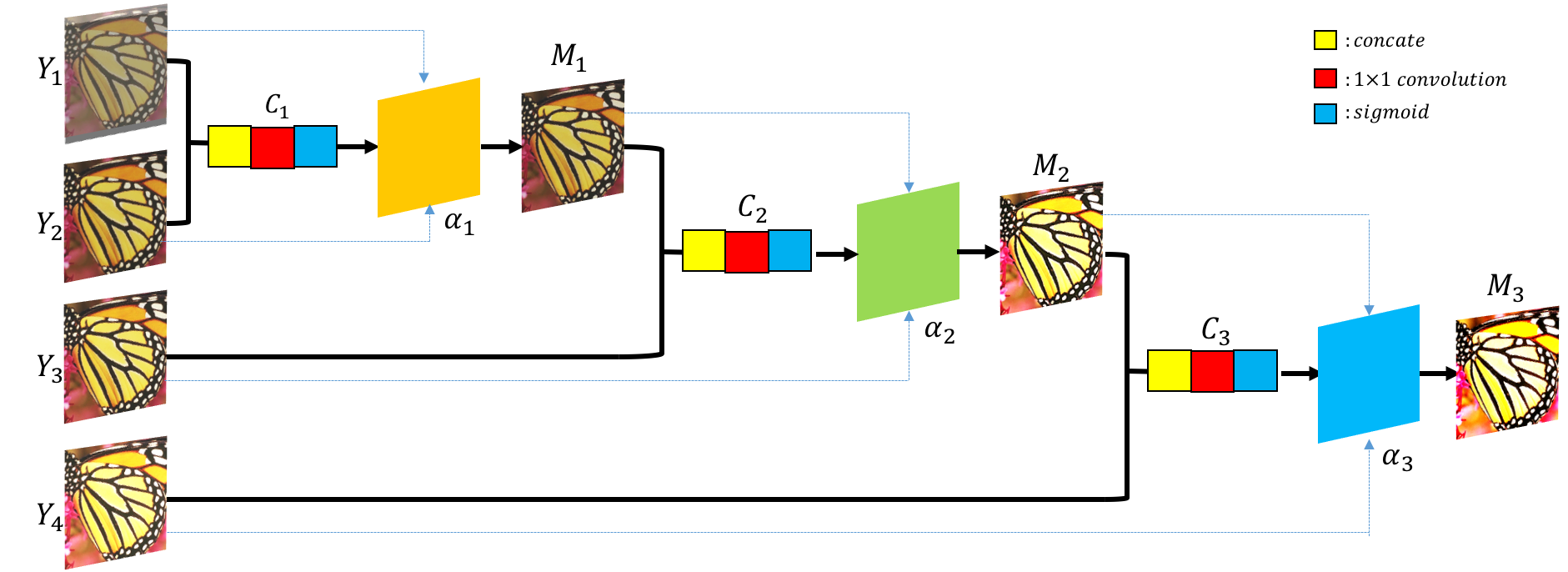}
\caption{\small A sketch of the adaptive element-wise fusion strategy, where $N=4$ and $M_{i}$ $(i=1,2,3)$ are the current fused outputs.}
\label{fig:F5}
\end{figure*}

\begin{algorithm}[ht]  
  \caption{Adaptive Element-wise Fusion Strategy.}  
  \begin{algorithmic}[1]
    \Require   
      Intermediate outputs $\{Y_{1},Y_{2},\dots,Y_{N} \}$.
    \Ensure  
      The final fused feature maps $M$. 
    \State Initialize $M$ with $Y_{1}$: $M=Y_{1}$;   
    \For{each $i\in [1,N-1]$}  
      \State Concatenate SR output $X=concat(M, Y_{i+1})$
      \State Convolve with 1$\times$1 tensor: $\alpha_{i} = C_{i}X$, $C_{i}$ is the $i$-th 1$\times$1 tensor
      \State Use sigmoid activation: $\alpha_{i} = sigmoid(\alpha_{i})$
      \State Update $M=\alpha_{i}M+(I-\alpha_{i})Y_{i+1}$
    \EndFor 
    \State return $M$ 
  \end{algorithmic}\label{alg1}  
\end{algorithm} 

Feature maps of different receptive fields are sensitive to features of different sizes, which are often fused to enhance the performance in various computer vision tasks. In our case, we develop an adaptive element-wise fusion strategy. With $N$ intermediate results $\{Y_{1},Y_{2},\dots,Y_{N} \}$ mapped from $\{F_{1},F_{2},\dots,F_{N} \}$ through U\&RNet, a list of weight tensors $\{W_{1},W_{2},\dots,W_{N}\}$ with the same size of output are determined by $\{Y_{1},Y_{2},\dots,Y_{N} \}$, which control how much of each raw result contributes to the final fused output. Here the adaptive weight tensors satisfy two traits:\\
Trait 1: Each adaptive tensor is determined by all intermediate outputs together, which can be formulated as
\begin{equation}
W_{i} = f_{i}(Y_{1},Y_{2},\dots,Y_{N}) , \ i = 1, 2, \dots, N,
\label{eq:trait1}
\end{equation}
where $f_{i}$ is the mapping from $\{Y_{1}, Y_{2}, \dots, Y_{N}\}$ to $W_{i}$;\\
Trait 2: The value of each point in the weight tensor is between 0 and 1, and
\begin{equation}
\sum_{i=1}^{N}W_{i}=I,
\label{eq:trait2}
\end{equation}
where $I$ is the tensor with all elements being 1. 

The final fused output $M$ is a convex weighted average of intermediate outputs $\{Y_{1},Y_{2},\dots,Y_{N} \}$: 
\begin{equation}
\begin{split}
I_{out}=M=\sum_{d=1}^{N}W_{d}Y_{d}.
\end{split}
\label{con:over_4}
\end{equation}

Inspired by the gate unit in LSTM, by adopting a series of $1\times1$ convolution kernels followed by sigmoid activation functions, we develop a heuristic algorithm to construct the fused output $M$, in which weight tensors satisfy the above two traits, as summarized in Algorithm~\ref{alg1}.
A sketch for Algorithm~\ref{alg1} is plotted in Fig.\ref{fig:F5}, in which intermediate variable tensor $\alpha_{i} \ (i=1,\ldots,N-1)$ is generated progressively given current SR outputs $\{Y_{1}, \dots, Y_{i}\}$, 1$\times$1 convolution kernel $C_{i}$ and sigmoid activation function. The use of sigmoid activation functions ensures the element-wise value of $\alpha_{i}$ to be between 0 and 1. The updating step (Step 6) ensures the output to be a convex weighted average of current inputs $\{Y_{1}, \dots, Y_{i}\}$.
From Algorithm~\ref{alg1}, $\{W_{1}, W_{2}, \dots,W_{N}\}$ can be obtained as
\begin{equation}
\begin{split}
W_{k}=\left\{
\begin{array}{lrc}
\displaystyle \prod_{i=1}^{N-1}\alpha_{i},   &      & {k = 1}; \\
\displaystyle \big (I-\alpha_{k-1} \big ) \big (\prod_{i=k}^{N-1}\alpha_{i} \big ),     &      & {2 \leq k < N}; \\
\displaystyle I-\alpha_{N-1},       &      & {k=N}, 
\end{array} \right.
\end{split}
\label{eq:wk}
\end{equation}
where $\alpha_{N-1}$ contains the information of $\{Y_{1}, Y_{2}, \dots, Y_{N}\}$. As (\ref{eq:wk}) shows that every $W_{k}$ contains $\alpha_{N-1}$, the first trait in (\ref{eq:trait1}) is satisfied; with simple algebra, the second trait in~(\ref{eq:trait2}) is also verified, and hence the rationality of the proposed methods.

\subsection{Loss Function for Training}

During training, we minimize the $\ell_{1}$ loss $L_{1}(x,y) = |x-y|$ over the training set of $M$ samples.
Let $X^{(i)}$ denote the $i$-th ground-truth HR label in the training set and $I_{out}^{(i)}$ denote the corresponding output of network. Then the loss function  $l$ is
\begin{equation}
l(I_{out}, X)=\frac{1}{M}\sum_{i=1}^{M}L_{1}(I_{out}^{(i)},X^{(i)}).
\end{equation}

When we apply the adaptive element-wise fusion strategy, we use the multi-supervised methods mentioned in~\cite{kim2016deeply,Tai-MemNet-2017} to train our model. The loss function of multi-supervised LCSCNet can be formulated as 
\begin{equation}
L(\Theta)=l(I_{out}, X) + \beta \sum_{d=1}^{N}l(Y_{d}, X),
\label{con:n1}
\end{equation}
where $I_{out}$ and $\{Y_{1},\dots, Y_{N} \}$ are defined in (\ref{con:over_4}), and $\beta$ is a trade-off parameter.

\section{Discussions}\label{s:s_4}

In this section, we mainly discuss the motivation and characteristics of Basic LCSCNet by showing its connections to DenseNet and its differences from ResNet and DenseNet. 

\subsection{Basic LCSCNet as an Efficient Variant of DenseNet}
\begin{figure*}
    \centering
    \subfloat[Original DenseBlock]{
       \includegraphics[scale=0.55]{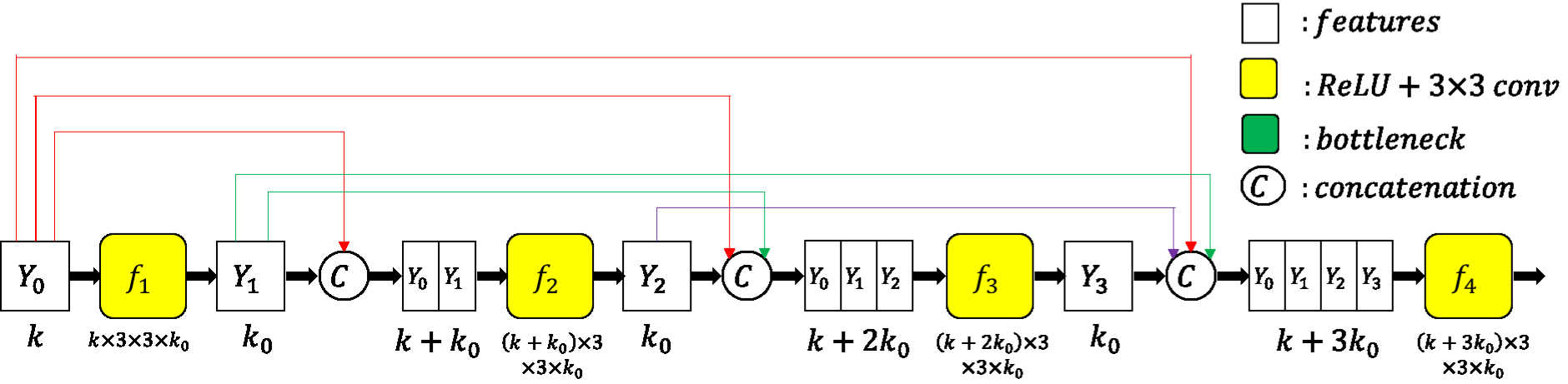}}
    \label{fig:D_1}\\
    \subfloat[DenseBlock with adjacent skip connection, equivalent to (a)]{
        \includegraphics[scale=0.38]{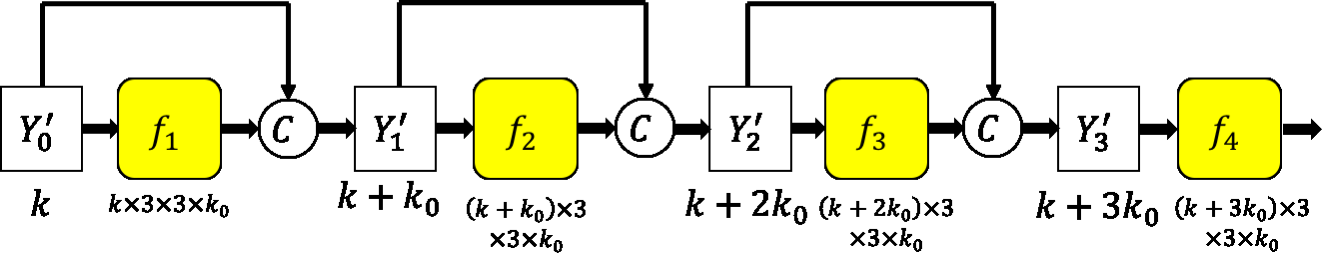}}
    \label{fig:D_2}\hfill
    \subfloat[DenseBlock with bottleneck (B-DenseBlock)]{
        \includegraphics[scale=0.3]{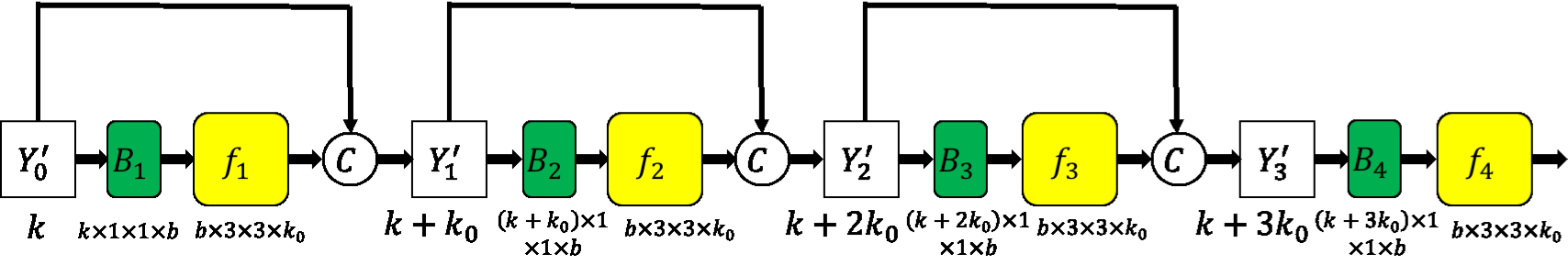}}
    \label{fig:D_3}\\
    \subfloat[Move forward every bottleneck layer]{
        \includegraphics[scale=0.3]{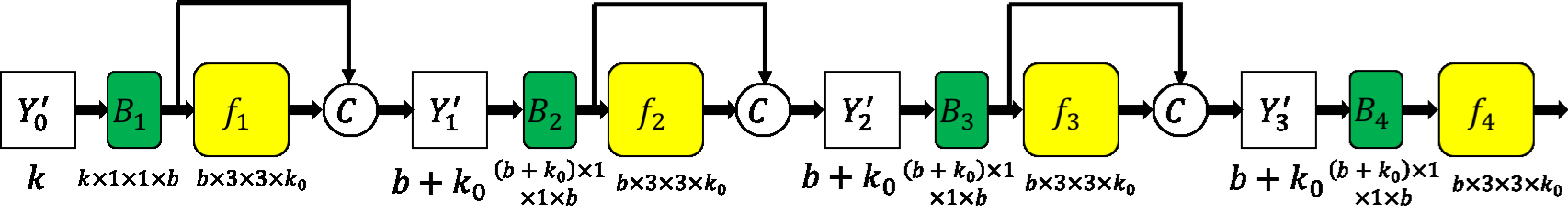}}
    \label{fig:D_4}\hfill
    \subfloat[Equivalent form of (d), Basic LCSCNet]{
        \includegraphics[scale=0.3]{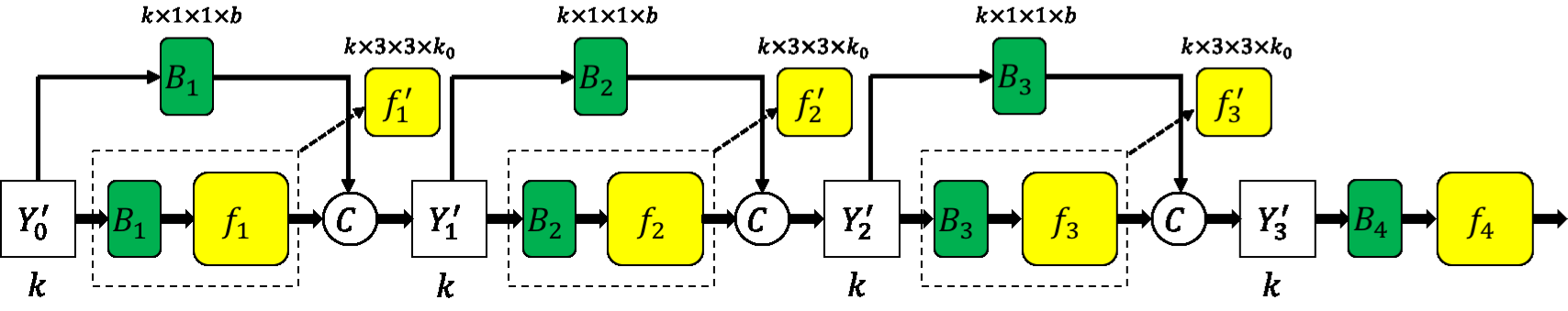}}
    \label{fig:D_5}\hfill
    \caption{\small Sketch on how a DenseBlock can be simplified into a Basic LCSCNet. For a better understanding, the channel number of each feature is marked beside the feature, and the kernel size of each convolution kernel is marked beside the kernel in form of ``input\_chanenel $\times$  kernel\_width $\times$ kernel\_height $\times$ output\_channel''.}
    \label{fig:D} 
\end{figure*}

In this sub-section, we illustrate that Basic LCSCNet can be transformed from DenseNet with small changes on topology: we first show the redundancy in DenseNet and then introduce Basic LCSCNet as a remedy for this redundancy.

Skip connection in DenseNet is implemented by directly concatenating all former features to be the input of current layer. For illustration, a 4-layer DenseBlock is depicted in Fig.\ref{fig:D}(a), in which $Y_{0}$ is a $k$-channel input feature, $Y_{i}$ is the newly-explored feature after nonlinear mapping $f_{i}$ ($i=1,2,3$), where $f_{i}$ consists of a ReLU followed by a $3\times 3$ convolution kernel with $k_{0}$ output channels ($k_{0}$ is also called growth rate in DenseNet). The last nonlinear mapping $f_{4}$ acts as a transition layer; $C$ means a concatenation operator in the channel dimension.

To have a better understanding of DenseBlock in Fig.\ref{fig:D}(a), we can simplify Fig.\ref{fig:D}(a) into its equivalent form in Fig.\ref{fig:D}(b), where $Y_{i}^{'}=concat(Y_{0},\dots,Y_{i})$ $(i=0,1,2,3)$. By denoting the concatenation of former features as $Y_{i}^{'}$, excessive skip connections in Fig.\ref{fig:D}(a) are simplified into concise adjacent skip connections. For simplicity, unless otherwise specified, we take the DenseBlock in the form of Fig.\ref{fig:D}(b) as the basic DenseBlock structure.

As shown in Fig.\ref{fig:D}(a) and Fig.\ref{fig:D}(b), when depth increases, the number of parameters of the convolution kernel in DenseNet also increases. To reduce the parameter amount, the authors of DenseNet applied the bottleneck layer\footnote{Many works add nonlinear activation before a $1\times 1$ convolution kernel to make the bottleneck layer; here we take the $1\times 1$ convolution kernel as the bottleneck layer.} before every nonlinear mapping and called it B-DenseNet. Fig.\ref{fig:D}(c) is the bottleneck version of Fig.\ref{fig:D}(b), where $B_{i}$ ($i=1,2,3$) is the $1\times 1$ convolution kernel with $b$ output channels. We can see that the parameter amount of every convolution kernel $f_{i}$ for nonlinear mapping is a constant now, and only the parameter amount of bottleneck layer with fewer parameters increases with depth. 

Although B-DenseBlock has reduced the parameter amount to a great extent, the parameter amount of each basic unit in B-DenseBlock still increases with depth. To further control the parameter amount, we make the parameter amount of each unit in B-DenseBlock a constant. One simple but effective solution is to move forward the bottleneck layer in each unit, reducing the number of channels of input feature to $b$ by the bottleneck layer before they are sent to the concatenation part, as shown in Fig.\ref{fig:D}(d). We can set $k=b+k_{0}$ to make channels of each feature unchanged. In this case, if we re-depict Fig.\ref{fig:D}(d) by allocating the bottleneck layer to each branch and using a nonlinear mapping $f_{i}^{'}$ to replace $B_{i} \circ f_{i}$, the structure of Basic LCSCNet reemerges, as shown in Fig.\ref{fig:D}(e). 

From the analysis above we can see that the $N$-layer Basic LCSCNet with an extra transition layer and the $(N+1)$-layer B-DenseNet share a strong relationship. This transition layer can be replaced by subsequent nonlinear operators and omitted. If it is replaced by a compressing layer located at the end of DenseBlock, it becomes BC-DenseNet. Since this compressing layer is to compress features generated by each block, when we simplify B-DenseUnit into LCSCUnit, the output channel of each LCSCUnit is already a constant, it is unnecessary to compress features again. From this perspective, BC-DenseNet can also be transferred to Basic LCSCNet in a similar way. Now look back into Fig.\ref{fig:F4}: the nonlinear output channel is just the growth rate in DenseNet, denoting how many new features are explored, and $1-\rho=\frac{n_{1}}{n_{1}+n_{2}}$ acts as some kind ``compress ratio'' denoting how many former features have flowed to the current stage through skip connections. 

\subsection{Differences from ResNet and DenseNet}
This sub-section aims to illustrate the differences between  Basic LCSCNet and ResNet/DenseNet as well as the novelty of our proposed network. It is still an open problem to compare different deep architectures. When different ways of skip connections are employed to alleviate training difficulties, the output features explored by nonlinear mapping with different skip connections have different constitutions. We suppose that by comparing different constitutions of the feature maps, we could get some useful information about the properties of different skip-connection architectures.

\subsubsection{Feature maps of ResNet} 

We use the structure in~\cite{he2016identity}. Let $Y_{k}$ and $Y_{k+1}$ denote the input and output of block $k$, respectively, and let $f_{k}^{R}(\cdot)$ denote the nonlinear transformation in block $k$. Then the mathematical formulation of block $k$ is
\begin{equation}
Y_{k+1}=Y_{k} + f_{k}^{R}(Y_{k}) =Y_{j} + \sum_{i=j}^{k}f_{i}^{R}(Y_{i}),\ 1 \leq j \leq k. 
\label{con:d2}
\end{equation}
From (\ref{con:d2}), we can see that in ResNet, skip connection is implemented by element-wise summation between adjacent features. Compared with traditional plain architecture, any former maps $Y_{j} \ (j=1,\ldots,k)$ can be added to the current state $Y_{k+1}$, creating many short paths for more ``smooth'' gradient flow during back-propagation. Moreover, it is extremely concise because no extra parameter is required for this skip connection. 

\subsubsection{Feature maps of DenseNet}

We employ the structure shown in Fig.\ref{fig:D}(b) to illustrate the properties of feature maps in DenseNet. Let $Y_{k}^{'}$ and $Y_{k+1}^{'}$ denote the input and output of unit $k$ in a DenseBlock, respectively, and $f_{k}^{D}(\cdot)$ the nonlinear transformation. Then the formulation of unit $k$ is 
\begin{equation}
\begin{split}
Y_{k+1}^{'} = concat(Y_{k}^{'}, f_{k}^{D}(Y_{k}^{'})),
\end{split}
\label{con:dense_1}
\end{equation}
and it follows that 
\begin{equation}
\begin{split}
Y_{k+1}^{'} = concat(Y_{j}^{'},f_{j}^{D}(Y_{j}^{'}),\dots,f_{k}^{D}(Y_{k}^{'})) ,~1 \leq j \leq k. 
\end{split}
\label{con:dense_2}
\end{equation}
Like ResNet, all the former features in DenseNet can be fused into the current stage, but instead of summation, all the feature maps are concatenated in the channel dimension. Such a skip connection has both advantages and disadvantages compared with ResNet. One obvious advantage is that when features produced in DenseNet are sent to follow-up convolution kernels to explore new features, the features from different stages use different convolution kernels, while in ResNet the reused parts and newly-explored ones share the same convolution kernel. From this perspective, connecting features by element-wise summation may restrict a network from reaching better solutions in some cases. As for disadvantage, concatenating features need more following convolution kernels. As shown in Fig.\ref{fig:D}(a) and Fig.\ref{fig:D}(b), the parameter amount of DenseUnit increases with depth. When a DenseNet is very deep, even a small growth rate may lead to a large parameter amount. 

\begin{table*}[htbp]
\centering
\caption{\small Quantitative comparisons on $\times 3$ SISR among the ResNet, B-DenseNet, BC-DenseNet and Basic LCSCNet of the same depth. {\color{blue}Blue} indicates the least parameters. {\color{red}Red} indicates the best quantitative performance.}
\label{chart:RDL}
\begin{tabular}{c|ccccc}
\hline
\thead{\textbf{Model}} & \thead{\textbf{Parameters}} & \thead{\textbf{Set5}} & \thead{\textbf{Set14}} & \thead{\textbf{BSD100}} & \thead{\textbf{Urban100}}\\
\hline
\thead{ResNet} & \thead{118.1K} & \thead{33.90/0.9233} & \thead{29.84/0.8328} & \thead{28.85/0.7987} & \thead{27.12/0.8303}\\
\hline
\thead{B-DenseNet} & \thead{219.8K} & \thead{33.98/0.9241} & \thead{29.87/{\color{red}0.8338}} & \thead{28.87/\color{red}0.7997} & \thead{\color{red}27.25/0.8326}\\
\hline
\thead{BC-Dense\_B3\_U10} & \thead{102.7K} & \thead{33.90/0.9234} & \thead{29.90/0.8336} & \thead{{\color{red}28.88}/0.7991} & \thead{27.22/0.8310}\\
\hline
\thead{BC-Dense\_B5\_U6} & \thead{90.4K} & \thead{33.92/0.9234} & \thead{{\color{red}29.90}/0.8334} & \thead{28.87/0.7990} & \thead{27.21/0.8307}\\
\hline 
\thead{Basic LCSCNet} & \thead{\color{blue}68.9K} & \thead{\color{red}33.99/0.9241} & \thead{29.87/0.8337} & \thead{28.87/0.7994} & \thead{27.24/0.8324}\\
\hline
\end{tabular}
\end{table*}

\subsubsection{Feature maps of Basic LCSCNet} 

From the analysis above, we can conclude that the feature re-usage of ResNet benefits from its concise skip connection between adjacent basic blocks and the new feature exploration of DenseNet mainly benefits from its little relevance between newly-explored feature maps and former ones. We have already seen that in Basic LCSCNet, former features are firstly compressed and then concatenated with the newly-explored features. Now we examine how the former features are combined in the current stage. Let $Y_{k}$ and $Y_{k+1}$ denote the input and output of the $k$-th LCSCUnit, and $K_{k}^{L}$ and $K_{k}^{NL}$ its convolution kernels. From Fig.\ref{fig:F4} and (\ref{eq:lcscunit}), we can derive the formulation of $1 \times 1$ convolution in the LC layer as   
\begin{align}
&Y_{k+1}^{L}(c_{o})
\nonumber \\                        
                   &=\sum_{c_{i}=1}^{n}K_{k}^{L}(c_{o},c_{i})Y_{k}(c_{i})  \nonumber \\
                   &=\sum_{c_{i}=1}^{n_{1}}K_{k}^{L}(c_{o},c_{i})Y_{k}^{L}(c_{i})+  
                   \sum_{c_{i}=n_{1}+1}^{n}K_{k}^{L}(c_{o},c_{i})Y_{k}^{NL}(c_{i}-n_{1})  \nonumber \\
                   &=\sum_{c_{i}=1}^{n_{1}}K_{k}^{L,L}(c_{o},c_{i})Y_{k}^{L}(c_{i})+
                   \sum_{c_{i}=1}^{n_{2}}K_{k}^{L,NL}(c_{o},c_{i})Y_{k}^{NL}(c_{i}), 
\label{con:d4}
\end{align}
where $c_{i}$ denotes the input channel and $c_{o}$ the output channel.

For simplicity, (\ref{con:d4}) can be rewritten as
\begin{equation}
\begin{split}
Y_{k+1}^{L}=K_{k}^{L,L}Y_{k}^{L}+K_{k}^{L,NL}Y_{k}^{NL}.
\end{split}
\label{con:d5}
\end{equation}    

Applying the same approach to the convolution kernel $K_{k}^{NL}$ in nonlinear transformation, we have
\begin{equation}
\begin{split}
Y_{k+1}^{NL}=K_{k}^{NL,L}ReLU(Y_{k}^{L})+K_{k}^{NL,NL}ReLU(Y_{k}^{NL}),
\end{split}
\label{con:d6}
\end{equation}
where $K_{k}^{NL,L}$ is the part of $K_{k}^{NL}$ only operating on $Y_{k}^{L}$ and $K_{k}^{NL,NL}$ only on $Y_{k}^{NL}$.

A `global' form of (\ref{con:d5}) is 
\begin{gather}
Y_{k+1}^{L}=P_{k+1}^{L}+P_{k+1}^{NL},\\
P_{k+1}^{L}=(\prod_{i=1}^{k}K_{i}^{L,L})Y_{1}^{L},\\ 
P_{k+1}^{NL}=\sum_{i=1}^{k-1}(K_{i}^{L,NL}\prod_{j=i+1}^{k}K_{j}^{L,L})Y_{i}^{NL}+K_{k}^{L,NL}Y_{k}^{NL}.
\label{con:d7}
\end{gather}

From (\ref{con:d5}), we can see that $Y_{k+1}^{L}$ restores the information of all former feature maps in the form of weighted summation. From (\ref{con:d6}), we can see that $Y_{k+1}^{NL}$ is the new features explored by new nonlinear transformation. Among the deep features produced by deep architectures, newly-explored parts are thought to be more important. In Basic LCSCNet, we concatenate newly-explored features with the former ones like DenseNet, ensuring that features of different kinds can be treated differently. Meanwhile, as former features in the current stage are mainly aimed to create paths for training deep networks, instead of concatenating each former features separately, we compress all the former features and then concatenate them with the newly-explored ones, making it quite parameter-economic like ResNet.

\section{Ablation Studies}\label{s:s_5}

\subsection{Comparison with ResNet and DenseNet}\label{s:s_5::A}

In this sub-section, we replace LCSCUnit in our basic LCSCNet by ResBlock or DenseUnit with the bottleneck layer. The three networks for comparison are all 34-layer, where Basic LCSCNet and DenseNet both have 30 units while ResNet has 15 blocks. As discussed before, the growth rate in DenseNet plays a similar role to the channel number of the nonlinear output. To compare fairly, if we set all output feature channels to 64 and $\rho$ of every LCSCUnit to 0.5, then the growth rate of DenseNet is 32 and the output channel of the bottleneck is 64. As for BC-DenseNet, for example, BC-Dense\_B3\_U10 means dividing the network into 3 blocks uniformly and add a compressing layer at the end of each block, whose output channel is 64. Here we train the above three models with the 291 dataset for $\times 3$ scale and the results are shown in Table~\ref{chart:RDL}. We use PSNR/SSIM~\cite{wang2004image} to measure reconstruction, and parameter amounts to measure storage efficiency. We can see that Basic LCSCNet has the least parameters and the competitive performance to DenseNet both better than ResNet.

\subsection{Efficiency Brought by the LC layer} \label{s:s_5::B}

\begin{table}[htbp]
\centering
\setlength{\tabcolsep}{1mm}{
\caption{\small Quantitative comparisons on $\times 3$ SISR between the original Basic LCSCNet and the Basic LCSCNet with $3 \times 3$ LC layers. {\color{red}Red} indicates the best quantitative performance.}
\label{chart:lc_layer}
\begin{tabular}{c|cccc}
\hline
\thead{\textbf{Model}} & \thead{\textbf{Set5}} & \thead{\textbf{Set14}} & \thead{\textbf{BSD100}} & \thead{\textbf{Urban100}}\\
\hline
\thead{Basic LCSCNet} & \thead{\color{red}33.99/0.9241} & \thead{29.87/\color{red}0.8337} & \thead{\color{red}28.87/0.7994} & \thead{\color{red}27.24/0.8324}\\
\hline
\thead{Basic LCSCNet \\ of $3 \times 3$ LC} & \thead{33.94/0.9238} & \thead{{\color{red}29.88}/0.8334} & \thead{28.87/0.7989} & \thead{27.17/0.8320} \\
\hline
\end{tabular}
}
\end{table}

Here we discuss the rationale behind implementing the LC layer with $1 \times 1$ convolution and its advantage on parameter efficiency. It is known that increasing receptive fields is essential for exploring deeper features. From Section~\ref{s:s_4} we can see that the LC layer helps transport the previous features and does not produce newly-explored features directly. Hence, we do not need to use $3 \times 3$ convolution to increase receptive fields in the LC layer and $1 \times 1$ convolution is sufficient. To support this view, we apply $3 \times 3$ convolution to the LC layer of the Basic LCSCNet mentioned in Section~\ref{s:s_5::A}. From Table~\ref{chart:lc_layer}, we can see that the LC layer with $3 \times 3$ convolution indeed does not achieve better performance.

The usage of $1 \times 1$ convolution as the LC layer also makes the proposed architecture more parameter-economic compared with ResNet and DenseNet. Firstly we compare Basic LCSCNet with ResNet. A basic unit in ResNet with $n_1$ input channels, $n_2$ output channels and a $k\times k$ nonlinear transformation convolution kernel has $n_1 n_2 k^2$ parameters. The number of parameters of a basic unit in Basic LCSCNet, with $n_1$ input channels, $n_2$ output channels, a $k\times k$ nonlinear transformation convolution kernel and parameter $\rho_0$, is $n_1n_2(k^2\rho_0+1-\rho_0)$. The ratio of parameter amounts of these two units with the same $n_1$, $n_2$ and $k$ is 
\begin{equation}
p_{L/R}(n_1,n_2,k)=\rho_0+\frac{1}{k^2}(1-\rho_0).
\label{con:L/R}
\end{equation}
As illustrated before, good performance can be obtained when $\rho_0$ is around 0.5. In practice, the size of a convolution kernel for feature extraction is usually an odd bigger than 3. So when $\rho_0$ is 0.5, $p_{L/R}(n_1,n_2,k)<55.7\%$, which means the parameter amount of Basic LCSCNet is just half of the ResNet's. 

As for DenseNet, the parameter amount of a basic unit increases with depth. We take B-DenseNet as example: if the output channel of nonlinear mapping in LCSCUnit and DenseUnit is both $n_{2}$, the output channel of $1\times 1$ compressing layer is both $n_{1}$, the nonlinear kernel size is $k\times k$, then the parameter amount of LCSCUnit is always  $(n_{1}+n_{2})(k^{2}n_{2}+n_{1})$, while the parameter amount of the $p$-th DenseUnit is $(pn_{2}n_{1}+k^{2}n_{1}n_{2})$. If such Basic LCSCNet and DenseNet both have $L$ nonlinear mapping layers, the ratio of parameter amounts of the two networks with the same $n_{1}$, $n_{2}$ and $k$ is 
\begin{equation}
\begin{split}
p_{L/D}(L;n_1,n_2,k)=\frac{2}{2k^{2}+L+1}(\frac{1}{\rho_0}+k^{2}\frac{1}{1-\rho_0}).
\end{split}
\label{con:L/D}
\end{equation}
From (\ref{con:L/D}), we can see the advantage of Basic LCSCNet is more remarkable when the network goes deeper. 
When we compare Basic LCSCNet with an $L$-layer BC-DenseNet of $N$ blocks, if the transition layer is omitted for simplicity, the ratio can be obtained by replacing $L$ with $\frac{L}{N}$ in (\ref{con:L/D}).

\subsection{Investigation into Parameter $\rho$}\label{s:s_5::C}

\begin{table*}[htbp]
\centering
\caption{\small Average $\times3$ PSNR/SSIM for Basic LCSCNets with different $\rho$ on the Set5, Set14, BSD100 and Urban100 datasets, respectively. {\color{red} Red} color indicates the best performance.}
\label{chart:C_1}
\begin{tabular}{c|ccccccc}
\hline
\thead{\boldsymbol{$\rho$}} & \thead{\textbf{0}} & \thead{\textbf{0.25}} & \thead{\textbf{0.375}} & \thead{\textbf{0.5}} & \thead{\textbf{0.625}} & \thead{\textbf{0.75}} & \thead{\textbf{1}} \\
\hline
\thead{Set5} & \thead{32.66/0.9103} & \thead{33.86/0.9229} & \thead{33.97/{\color{red} 0.9242}} & \thead{{\color{red} 33.99}/0.9241} & \thead{33.94/0.9241} & \thead{33.92/0.9237} & \thead{31.78/0.8941}\\
\hline
\thead{Set14} & \thead{29.27/0.8208} & \thead{29.82/0.8330} & \thead{29.90/0.8337} & \thead{29.87/0.837} & \thead {\color{red} {29.93/0.8340}} &\thead{29.85/0.8333} & \thead{28.57/0.8012}\\
\hline
\thead{BSD100} & \thead{28.41/0.7858} & \thead{28.85/0.7984} & \thead{\color{red}{28.88/0.7997}} & \thead{28.87/0.7994} & \thead{28.87/0.7994} & \thead{28.85/0.7990} & \thead{27.92/0.7648}\\
\hline
\thead{Urban100} & \thead{26.21/0.8011} & \thead{27.16/0.8296} & \thead{\color{red}{27.25/0.8329}} & \thead{27.24/0.8324} & \thead{27.24/0.8321} & \thead{27.20/0.8312} & \thead{25.50/0.7761}\\
\hline
\end{tabular}
\end{table*}

\subsubsection{Fixed $\rho$ throughout the network}

In this situation, we find when $\rho$ is around 0.5, the best performance could be achieved. Table~\ref{chart:C_1} shows 34-layer Basic LCSCNets for $\times$3 scale with different fixed $\rho$. As here we mainly focus on the effect of $\rho$, the experiments are conducted without adaptive element-wise fusion. Firstly, we consider two special cases of $\rho$. When $\rho$ is 0, the feature exploration part is a linear transformation; if the upsampling and reconstruction part is taken into account, the whole network has just two nonlinear convolution layers, whose fitting capacity for complex functions is relatively poor. In contrast, when $\rho$ is 1, Basic LCSCNet becomes the traditional feedforward neural network without skip connections, which is difficult to train. Hence $\rho$ balances the fitting capacity and the training ease of Basic LCSCNet. As Table~\ref{chart:C_1} shows, when $\rho=0.25$, the performance is suboptimal because of the restricted fitting capacity; when $\rho=0.75$, the performance is suboptimal mainly because the LC layers output fewer feature maps. As we discussed before, the output feature maps of LC layers restore the information of former features, insufficiency of which leads to insufficient skip connections and thus training difficulty increase and performance decline.

\begin{table}[htbp]
\centering
\setlength{\tabcolsep}{1mm}{
\caption{\small The effect of ordinal position of block with different $\rho$ on average $\times3$ PSNR/SSIM for the Set5, Set14, BSD100 and Urban100 datasets. Each block has the same number of LCSCUnits. }
\label{chart:C_2}
\begin{tabular}{c|cc|cc}
\hline
\thead{\boldsymbol{$\rho$} \textbf{list}} & \thead{\textbf{[0.5,0.75]}} & \thead{\textbf{[0.75,0.5]}} & \thead{\textbf{[0.5,0.625,0.75]}} & \thead{\textbf{[0.75,0.625,0.5]}}  \\
\hline
\thead{Set5} & \thead{33.97/0.9240} & \thead{34.02/0.9244} & \thead{33.89/0.9234} & \thead{33.95/0.9239}  \\
\hline
\thead{Set14} & \thead{29.91/0.8341} & \thead{29.91/0.8343} & \thead{29.86/0.8332} & \thead{29.86/0.8336} \\
\hline
\thead{BSD100} & \thead{28.88/0.7998} & \thead{28.89/0.8001} & \thead{28.86/0.7993} & \thead{28.88/0.7994} \\
\hline
\thead{Urban100} & \thead{27.28/0.8336} & \thead{27.31/0.8343} & \thead{27.20/0.8317} & \thead{27.25/0.8323} \\
\hline
\end{tabular}}
\end{table}

\subsubsection{Different $\rho$ throughout the network}   

In this situation, the LCSCUnits with the same $\rho$ form an LCSCBlock and different LCSCBlocks have different $\rho$s. We find that as depth increases, the $\rho$ of an LCSCBlock should be decreased slightly to improve performance. Table~\ref{chart:C_2} shows the relevant experimental results on the 34-layer Basic LCSCNets with different ordinal positions of $\rho$ list for $\times$3 scale. One possible reason for this phenomenon is that as depth increases, exploring higher-level features becomes harder, so there is less room for newly-explored features. Meanwhile, as information on former feature maps accumulates, more room is needed for reusing former features.

\subsection{Ablation Studies on Different Fusion Strategies} \label{s:s_5::D}

Table~\ref{chart:fusion_trait} compares properties of the vectorized fusion~\cite{kim2016deeply}, MSCN~\cite{liu2016learning} and our proposed fusion method. We can see that our method incorporates the advantages of the vectorized fusion and MSCN. We also compare these fusion strategies quantitatively. We train 34-layer LCSCNets for $\times$2 scale with $\rho$ list [0.75, 0.6875, 0.625, 0.5625, 0.5], and every six LCSCUnits with the same $\rho$ form a LCSCBlock. In Table~\ref{chart:C_3}, Basic LCSCNet, LCSCNet\_S, LCSCNet\_M and LCSCNet denote the LCSCNet without any fusion, with vectorized fusion, with MSCN and with our proposed method, respectively. Here we note that our implementation of MSCN is slightly different from the original one. In the original MSCN, the input of the weight module is the bicubic of LR, while in our LCSCNet we use the upsampled LR input. This small difference should have little influence on final results. As Table~\ref{chart:C_3} shows, when combined with Basic LCSCNet, our fusion strategy performs better than the other two fusion benchmarks. 

\begin{table}[htbp]
\centering
\setlength{\tabcolsep}{1mm}{
\caption{\small Brief comparisons among different fusion strategies.}
\label{chart:fusion_trait}
\begin{tabular}{c|c|c|c}
\hline
\thead{} & \thead{\footnotesize{Vectorized~fusion\cite{kim2016deeply}}} & \thead{\footnotesize{MSCN\cite{liu2016learning}}} & \thead{\footnotesize{Our~method}} \\
\hline
\thead{\footnotesize Adaptiveness} & \thead{$\times$} & \thead{$\surd$} & \thead{$\surd$} \\
\hline
\thead{\footnotesize Pixel-wise} & \thead{$\times$} & \thead{$\surd$} & \thead{$\surd$} \\ 
\hline
\thead{\footnotesize Normalization} & \thead{$\surd$} & \thead{$\times$} & \thead{$\surd$} \\
\hline
\thead{\footnotesize Multi-supervised \\ training} & \thead{$\surd$} & \thead{$\times$} & \thead{$\surd$} \\
\hline
\end{tabular}
}
\end{table}

\begin{table}[htbp]
\centering
\setlength{\tabcolsep}{1mm}{
\caption{\small Average $\times2$ PSNR/SSIM for LCSCNets with different fusions on the Set5, Set14, BSD100 and Urban100 datasets, respectively. {\color{red} {Red}} indicates the best results. }
\label{chart:C_3}
\begin{tabular}{c|cccc}
\hline
\thead{} & \thead{\textbf{Basic LCSCNet}} & \thead{\textbf{LCSCNet\_S}} & \thead{\textbf{LCSCNet\_M}} & \thead{\textbf{LCSCNet}} \\
\hline
\thead{Set5} & \thead{37.77/0.0.9558} & \thead{37.80/\color{red} 0.9560} & \thead{37.79/0.9559} & \thead{{\color{red} 37.84}/0.9559} \\
\hline
\thead{Set14} & \thead{33.23/0.9140} & \thead{33.26/0.9144} & \thead{33.25/0.9142} & \thead{\color{red} 33.31/0.9144} \\
\hline
\thead{BSD100} & \thead{32.06/0.8980} & \thead{32.05/0.8981} & \thead{32.07/0.8981} & \thead{\color{red} 32.08/0.8984} \\
\hline
\thead{Urban100} & \thead{31.15/0.9182} & \thead{31.26/0.9197} & \thead{31.23/0.9190} & \thead{\color{red} 31.31/0.9200} \\
\hline
\end{tabular}}
\end{table}

\subsection{Ablation Studies on LCSCBlock and E-LCSCBlock} \label{s:s_5::E}

\begin{figure}[htbp]
\centering
\includegraphics[scale=0.5]{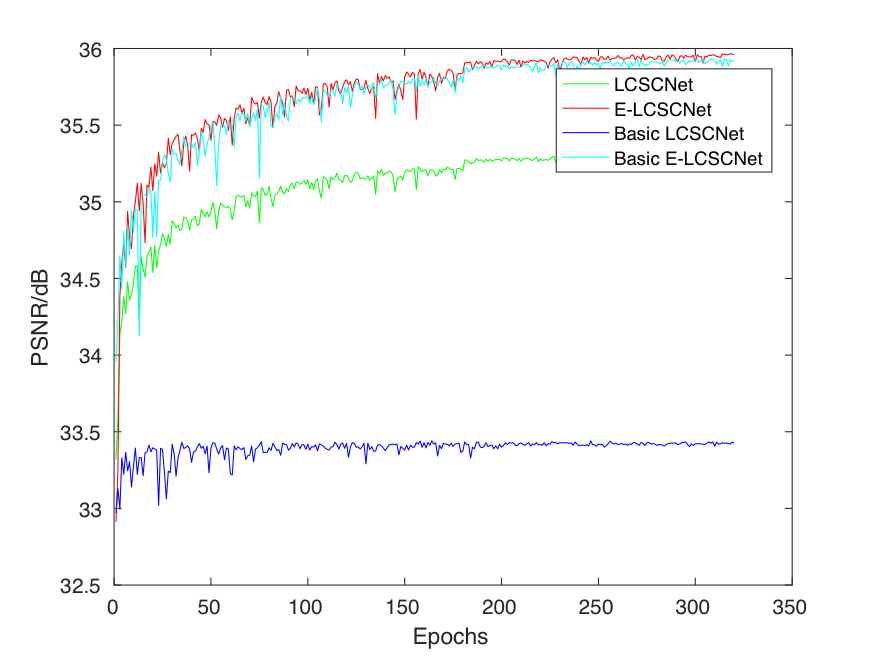}
\caption{\small Convergence comparison between deep wide (Basic) LCSCNet and (Basic) E-LCSCNet on the DIV2K validation set for scale $\times 2$.}
\label{fig:elcsc}
\end{figure} 

Firstly, we show that when we use LCSCUnits to construct deep models of moderate scales, the advantage of E-LCSCBlock is mild. With the 291 dataset, we train 34-layer, 44-layer and 54-layer Basic LCSCNets for $\times 3$ scale, $\rho$ of each unit is 0.5, and the number of feature channels is 64. For comparison, we use 10 LCSCUnits to constitute an E-LCSCBlock and train 37-layer, 48-layer and 59-layer Basic E-LCSCNets, respectively. As Table~\ref{chart:EandL} shows, every Basic E-LCSCNet performs  better than its corresponding Basic LCSCNet. 

\begin{table*}
    \centering
    \caption{\small Average $\times 3$ PSNR/SSIM for Basic LCSCNet and its corresponding Basic E-LCSCNet on Set5, Set14, BSD100 and Urban100. All the models are of moderate scales (Parameter amount $<$ 150K).}
    \label{chart:EandL}
    \begin{tabular}{c|cc|cc|cc}
         \hline
         \thead{} & \thead{LC\_34} & \thead{E-LC\_37} & \thead{LC\_44} & \thead{E-LC\_48} & \thead{LC\_54} & \thead{E-LC\_59} \\
         \hline
         \thead{Set5} & \thead{33.99/0.9241} & \thead{34.01/0.9248} & \thead{34.02/0.9244} & \thead{34.05/0.9251} & \thead{34.03/0.9244} & \thead{34.08/0.9248} \\
         \hline
         \thead{Set14} & \thead{29.87/0.8337} & \thead{29.92/0.8349} & \thead{29.85/0.8334} & \thead{29.90/0.8345} & \thead{29.88/0.8340} & \thead{29.89/0.8339} \\
         \hline
         \thead{BSD100} & \thead{28.87/0.7994} & \thead{28.89/0.8002} & \thead{28.87/0.7996} & \thead{28.90/0.8004} & \thead{28.89/0.7998} & \thead{28.89/0.7998} \\
         \hline
         \thead{Urban100} & \thead{27.24/0.8324} & \thead{27.28/0.8340} & \thead{27.23/0.8326} & \thead{27.29/0.8343} & \thead{27.27/0.8330} & \thead{27.32/0.8347} \\
         \hline
    \end{tabular}
\end{table*}

Then we show that when we aim to develop an extremely deep and wide network, E-LCSCBlock can make up the deficiencies of LCSCBlock. With the DIV2K dataset we train a Basic LCSCNet for $\times 2$ scale of $\rho$ list [0.75, 0.71875, 0.6875, 0.65625, 0.625, 0.59375, 0.5625, 0.53125, 0.5], every sixteen LCSCUnits with the same $\rho$ form a LCSCBlock, and the output channel of each feature is 128. Its convergence curve is the blue one in Fig.\ref{fig:elcsc}, indicating quite poor performance. For comparison, we train the LCSCNet with the same setting, and its performance (the green curve in Fig.\ref{fig:elcsc}) is significantly better than Basic LCSCNet. We contribute this obvious improvement to the extra short paths created by the adaptive fusion strategy, which suggests that more short paths may further help in this case. The experimental results shown in Fig.\ref{fig:elcsc} also support this view: when we evolve (Basic) LCSCNet into (Basic) E-LCSCNet, the performance of deep architecture booms.

\section{Experimental Results}\label{s:s_6}

\subsection{Comparison with State-of-the-Art Models}

It is well known that the training set and the parameter amount largely influence the final performance of a model. To compare with various representative models fairly, we divide these models into three categories: models trained on the 291 dataset~\cite{yang2010image,martin2001database}, light models (Params $<$ 2M) trained on the DIV2K dataset~\cite{Agustsson_2017_CVPR_Workshops} and large models (Params $>$ 10M) on DIV2K. When compared with models on the 291 dataset such as VDSR~\cite{kim2016accurate}, DRCN~\cite{kim2016deeply}, LapSR~\cite{LapSRN}, DRRN~\cite{tai2017image} and MemNet~\cite{Tai-MemNet-2017}, we train a 76-layer LCSCNet with the proposed fusion strategy, $\rho$ list is also [0.75, 0.71875, 0.6875, 0.65625, 0.625, 0.59375, 0.5625, 0.53125, 0.5] but every eight units with the same $\rho$ form a block, denoted by LCSC\_76\_291. When compared with light models on DIV2K and similarly large datasets such as SelNet~\cite{choi2017deep}, SRDenseNet~\cite{tong2017image}, CARN~\cite{ahn2018fast} and FALSR-A~\cite{chu2019fast}, because the fusion part is quite computation-consuming, our light models was developed just based on Basic E-LCSCNet. Our light models share the same $\rho$ list with LCSC\_76\_291, but every six units with the same $\rho$ form a block, denoted by BE-LCSC\_L. When compared with large models on DIV2K such as EDSR~\cite{lim2017enhanced} and RDN~\cite{zhang2018residual}, the E-LCSCNet mentioned in Section~\ref{s:s_5::E} is adopted. 

\begin{table*}[htbp]
\centering
\footnotesize
\caption{\footnotesize Quantitative comparisons among mainstream deep models for SISR. To compare fairly, we divide models into three categories: models trained on 291, light models (Params $<$ 2M) trained on DIV2K, and large models trained on DIV2K. For each scale, we compare the models within the same category, and the best performance is highlighted in {\color{red} Red}. In DRCN, MemNet, LCSC\_76\_291 and E-LCSCNet, extra Mult\&Adds of the multi-supervised fusion part are added after the Mult\&Adds of the basic structure.}
\label{chart:big}
\begin{tabular}{c|cccccccc}
\hline
\thead{\textbf{Scale}} & \thead{\textbf{Model}} & \thead{\textbf{Training data}} & \thead{\textbf{Params}} & \thead{\textbf{Mult\&Adds}} & \thead{\textbf{Set5}} & \thead{\textbf{Set14}} & \thead{\textbf{BSD100}} & \thead{\textbf{Urban100}} \\
\hline
\multirow{12}{*}{\thead{$\times 2$}} 
        & \thead{\thead{VDSR} \\ \thead{DRCN} \\ \thead{LapSRN} \\ \thead{DRRN} \\ \thead{MemNet} \\ \thead{LCSC\_76\_291}}
        & \thead{\thead{291} \\ \thead{291} \\ \thead{291} \\ \thead{291} \\ \thead{291} \\ \thead{291}}
        & \thead{\thead{665K} \\ \thead{1774K} \\ \thead{813K} \\ \thead{297K} \\ \thead{667K}\\ \thead{1844K}}
        & \thead{\thead{612.6G} \\ \thead{9243.0G+8731.3G} \\ \thead{29.9G} \\ \thead{6796.9G} \\ \thead{2261.8G+3.2G}  \\ \thead{407.8G+616.3G}}
        & \thead{\thead{37.53/0.9587} \\ \thead{37.63/0.9588} \\ \thead{37.52/0.9591} \\ \thead{37.74/0.9591} \\ \thead{37.78/0.9597} \\ \thead{\color{red}37.86/0.9600}}
        & \thead{\thead{33.03/0.9124} \\ \thead{33.04/0.9118} \\ \thead{33.08/0.9130} \\ \thead{33.23/0.9136} \\ \thead{33.28/0.9142} \\ \thead{\color{red}33.34/0.9146}}
        & \thead{\thead{31.90/0.8960} \\ \thead{31.85/0.8942} \\ \thead{31.80/0.8950} \\ \thead{32.05/0.8973} \\ \thead{32.08/0.8978} \\ \thead{\color{red}32.10/0.8985}}
        & \thead{\thead{30.76/0.9140} \\ \thead{30.75/0.9133} \\ \thead{30.41/0.9101} \\ \thead{31.23/0.9188} \\ \thead{31.31/0.9195} \\ \thead{\color{red}31.34/0.9204}} \\
        \cline{2-9}
        & \thead{\thead{SelNet} \\ \thead{CARN} \\ \thead{FALSR-A} \\  \thead{BE-LCSC\_L}}
        & \thead{\thead{DIV2K} \\ \thead{DIV2K} \\ \thead{DIV2K} \\ \thead{DIV2K}}
        & \thead{\thead{974K} \\ \thead{1592K} \\ \thead{1021K} \\ \thead{1552K}}
        & \thead{\thead{225.7G} \\ \thead{222.8G} \\ \thead{234.7G} \\ \thead{358.6G}}
        & \thead{\thead{37.89/0.9598} \\ \thead{37.76/0.9590} \\ \thead{37.82/0.9595} \\ \thead{\color{red}38.01/0.9600}}
        & \thead{\thead{33.61/0.9160} \\ \thead{33.52/0.9166} \\ \thead{33.55/\color{red}0.9168} \\  \thead{{\color{red}33.67}/0.9160}}
        & \thead{\thead{32.08/0.8984} \\ \thead{32.09/0.8978} \\ \thead{32.12/0.8987} \\ \thead{\color{red}32.23/0.9002}}
        & \thead{\thead{-/-} \\ \thead{31.92/0.9256}\\ \thead{31.93/0.9256} \\ \thead{\color{red}32.31/0.9297}} \\
        \cline{2-9}
        & \thead{\thead{EDSR} \\ \thead{D\_DBPN} \\ \thead{RDN} \\
        \thead{E-LCSCNet}}
        & \thead{\thead{DIV2K} \\ \thead{DIV2K+Flickr} \\ \thead{DIV2K} \\
        \thead{DIV2K}}
        & \thead{\thead{40.7M} \\ \thead{5876.3K} \\ \thead{22.1M} \\ \thead{14.2M}}
        & \thead{\thead{9379.4G} \\ \thead{3429.0G} \\ \thead{5096.2G} \\ \thead{3126.4G+1251.7G}}
        & \thead{\thead{38.11/0.9602} \\ \thead{38.09/0.9600} \\ \thead{\color{red}38.24/0.9614} \\ \thead{38.23/0.9608}}
        & \thead{\thead{33.92/0.9195} \\ \thead{33.87/0.9191} \\ \thead{\color{red}34.01/0.9212} \\ \thead{33.85/0.9180}}
        & \thead{\thead{32.32/0.9013} \\ \thead{32.27/0.9000} \\ \thead{32.34/0.9017} \\ \thead{\color{red}32.36/0/9018}}
        & \thead{\thead{32.93/0.9351} \\ \thead{32.55/0.9324} \\ \thead{32.89/\color{red}{0.9353}} \\ \thead{{\color{red}32.93}/0.9351}} \\
\hline
\multirow{12}{*}{\thead{$\times 3$}} 
        & \thead{\thead{VDSR} \\ \thead{DRCN} \\ \thead{LapSRN} \\ \thead{DRRN} \\ \thead{MemNet} \\ \thead{LCSC\_76\_291}}
        & \thead{\thead{291} \\ \thead{291} \\ \thead{291} \\ \thead{291} \\ \thead{291} \\ \thead{291}}
        & \thead{\thead{665K} \\ \thead{1774K} \\ \thead{813K} \\ \thead{297K} \\ \thead{667K}\\ \thead{1844K}}
        & \thead{\thead{612.6G} \\ \thead{9243.0G+8731.3G} \\ \thead{29.9G} \\ \thead{6796.9G} \\ \thead{2261.8G+3.2G}  \\ \thead{181.3G+616.3G}}
        & \thead{\thead{33.66/0.9213} \\ \thead{33.82/0.9226} \\ \thead{33.82/0.9227} \\ \thead{34.03/0.9244} \\ \thead{34.09/0.9248} \\ \thead{\color{red}34.13/0.9254}}
        & \thead{\thead{29.77/0.8314} \\ \thead{29.76/0.8311} \\ \thead{29.79/0.8320} \\ \thead{29.96/0.8349} \\ \thead{\color{red}30.00/0.8350} \\ \thead{29.95/0.8348}}
        & \thead{\thead{28.82/0.7976} \\ \thead{28.80/0.7963} \\ \thead{28.82/0.7973} \\ \thead{28.95/0.8004} \\ \thead{28.96/0.8001} \\ \thead{\color{red}28.97/0.8014}}
        & \thead{\thead{27.14/0.8279} \\ \thead{27.15/0.8276} \\ \thead{27.07/0.8272} \\ \thead{27.53/{\color{red}0.8378}} \\ \thead{{\color{red}27.56}/0.8376} \\ \thead{27.53/0.8377}} \\
        \cline{2-9}
        & \thead{\thead{SelNet} \\ \thead{CARN} \\  \thead{BE-LCSC\_L}}
        & \thead{\thead{DIV2K} \\ \thead{DIV2K} \\ \thead{DIV2K}}
        & \thead{\thead{1159K} \\ \thead{1592K} \\ \thead{1736K}}
        & \thead{\thead{120.0G} \\ \thead{118.8G} \\ \thead{179.1G}}
        & \thead{\thead{34.27/0.9257} \\ \thead{34.29/0.9255} \\ \thead{\color{red}34.39/0.9265}}
        & \thead{\thead{30.30/0.8399} \\ \thead{30.29/{\color{red}0.8407}} \\ \thead{{\color{red}30.33}/0.8395}}
        & \thead{\thead{28.97/0.8025} \\ \thead{29.06/0.8034} \\ \thead{\color{red}29.12/0.8065}}
        & \thead{\thead{-/-} \\ \thead{28.06/0.8493} \\ \thead{\color{red}28.25/0.8540}} \\
        \cline{2-9}
        & \thead{\thead{EDSR} \\ \thead{D\_DBPN} \\ \thead{RDN} \\
        \thead{E-LCSCNet}}
        & \thead{\thead{DIV2K} \\ \thead{DIV2K+Flickr} \\ \thead{DIV2K} \\
        \thead{DIV2K}}
        & \thead{\thead{43.7M} \\ \thead{-} \\ \thead{22.3M} \\
        \thead{14.9M}}
        & \thead{\thead{4471.8G} \\ \thead{-} \\ \thead{2284.7G} \\
        \thead{1389.5G+1251.7G}}
        & \thead{\thead{34.65/0.9280} \\ \thead{-/-} \\ \thead{34.71/\color{red}0.9296} \\ \thead{{\color{red}34.71}/0.9286}}
        & \thead{\thead{30.52/0.8462} \\ \thead{-/-} \\ \thead{\color{red}30.57/0.8468} \\ \thead{30.56/0.8460}} 
        & \thead{\thead{29.25/0.8093} \\ \thead{-/-} \\ \thead{29.26/0.8093} \\ \thead{\color{red}29.27/0.8104}} 
        & \thead{\thead{28.80/0.8653} \\ \thead{-/-} \\ \thead{28.80/0.8653} \\ \thead{\color{red}28.83/0.8658}} \\
\hline
\multirow{12}{*}{\thead{$\times 4$}} 
        & \thead{\thead{VDSR} \\ \thead{DRCN} \\ \thead{LapSRN} \\ \thead{DRRN} \\ \thead{MemNet} \\ \thead{LCSC\_76\_291}}
        & \thead{\thead{291} \\ \thead{291} \\ \thead{291} \\ \thead{291} \\ \thead{291} \\ \thead{291}}
        & \thead{\thead{665K} \\ \thead{1774K} \\ \thead{813K} \\ \thead{297K} \\ \thead{667K}\\ \thead{1844K}}
        & \thead{\thead{612.6G} \\ \thead{9243.0G+8731.3G} \\ \thead{29.9G} \\ \thead{6796.9G} \\ \thead{2261.8G+3.2G}  \\ \thead{110.0G+616.3G}}
        & \thead{\thead{31.35/0.8838} \\ \thead{31.53/0.8854} \\ \thead{31.54/0.8855} \\ \thead{31.68/0.8888} \\ \thead{31.74/0.8893} \\ \thead{\color{red}31.76/0.8899}}
        & \thead{\thead{28.01/0.7674} \\ \thead{28.02/0.7670} \\ \thead{28.19/0.7720} \\ \thead{28.21/0.7721} \\ \thead{{\color{red}28.26}/0.7723} \\ \thead{28.20/{\color{red}0.7731}}}
        & \thead{\thead{27.29/0.7251} \\ \thead{27.23/0.7233} \\ \thead{27.32/0.7280} \\ \thead{27.38/0.7284} \\ \thead{{\color{red}27.40}/0.7281} \\ \thead{27.36/{\color{red}0.7293}}}
        & \thead{\thead{25.18/0.7524} \\ \thead{25.14/0.7510} \\ \thead{25.21/0.7553} \\ \thead{25.44/0.7638} \\ \thead{{\color{red}25.50}/0.7638} \\ \thead{25.38/\color{red}0.7643}} \\
        \cline{2-9}
        & \thead{\thead{SRDenseNet} \\  \thead{SelNet} \\ \thead{CARN} \\  \thead{BE-LCSC\_L}}
        & \thead{\thead{ImageNet Subset} \\ \thead{DIV2K} \\ \thead{DIV2K} \\ \thead{DIV2K}}
        & \thead{\thead{2015K} \\ \thead{1417K} \\ \thead{1592K} \\ \thead{1699K}}
        & \thead{\thead{389.9G} \\ \thead{83.1G} \\ \thead{90.9G} \\ \thead{124.8G}}
        & \thead{\thead{32.02/0.8934} \\ \thead{32.00/0.8931} \\ \thead{32.13/0.8937} \\ \thead{\color{red}32.20/0.8948}}
        & \thead{\thead{28.50/0.7782} \\ \thead{28.49/0.7783} \\ \thead{28.60/0.7806} \\ \thead{\color{red}28.66/0.7806}}
        & \thead{\thead{27.53/0.7337} \\ \thead{27.44/0.7325} \\ \thead{27.58/0.7349} \\ \thead{\color{red}27.62/0.7390}}
        & \thead{\thead{26.05/0.7819} \\ \thead{-/-} \\ \thead{26.07/0.7837}  \\ \thead{\color{red}26.22/0.7908}} \\
        \cline{2-9}
        & \thead{\thead{EDSR} \\ \thead{D\_DBPN} \\ \thead{RDN} \\
        \thead{E-LCSCNet}}
        & \thead{\thead{DIV2K} \\ \thead{DIV2K+Flickr} \\ \thead{DIV2K} \\ \thead{DIV2K}}
        & \thead{\thead{43.1M} \\ \thead{10.3M} \\ \thead{22.6M} \\ \thead{14.8M}}
        & \thead{\thead{2890.0G} \\ \thead{5715.4G} \\ \thead{1300.7G} \\ \thead{781.6G+1700.7G}}
        & \thead{\thead{32.46/0.8968} \\ \thead{32.47/0.8980} \\ \thead{32.47/{\color{red}0.8990}} \\ \thead{{\color{red}32.51}/0.8984}}
        & \thead{\thead{28.80/{\color{red}0.7876}} \\ \thead{{\color{red}28.82}/0.7860} \\ \thead{28.81/0.7871} \\ \thead{28.81/0/7871}}
        & \thead{\thead{27.71/0.7420} \\ \thead{27.72/0.7400} \\ \thead{27.72/0.7419} \\ \thead{\color{red}27.73/0.7433}}
        & \thead{\thead{26.64/0.8033} \\ \thead{26.38/0.7946} \\ \thead{26.61/0.8028} \\ \thead{\color{red}26.64/0.8033}} \\
\hline
\end{tabular}
\end{table*}

\begin{figure}[htbp]
    \centering
    \subfloat[\scriptsize{HR}]{
       \includegraphics[scale=0.29]{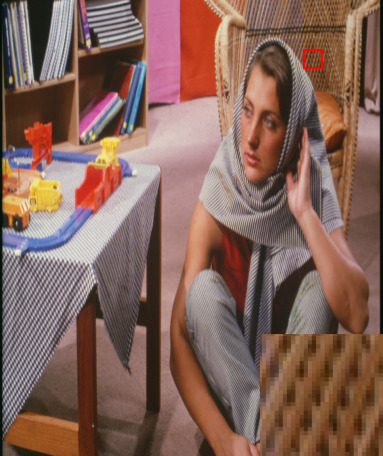}}
    \label{fig:barbara_HR}\hfill
    \subfloat[\scriptsize{VDSR}]{
        \includegraphics[scale=0.29]{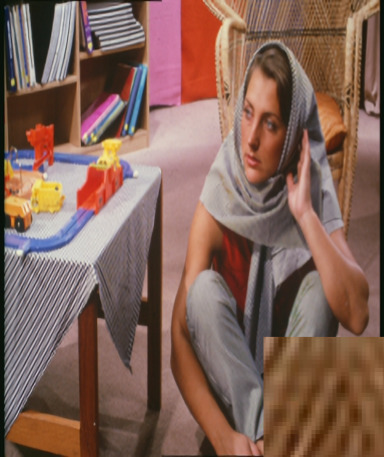}}
    \label{fig:barbara_VDSR}\hfill
    \subfloat[\scriptsize{DRCN}]{
        \includegraphics[scale=0.29]{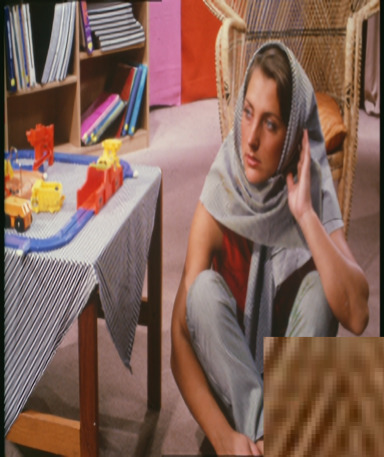}}
    \label{fig:barbara_DRCN}\hfill
    \subfloat[\scriptsize{LapSR}]{
        \includegraphics[scale=0.29]{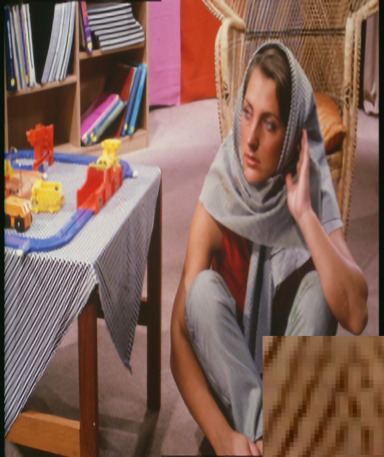}}
    \label{fig:barbara_LapSR} \\
    \subfloat[\scriptsize{DRRN}]{
        \includegraphics[scale=0.29]{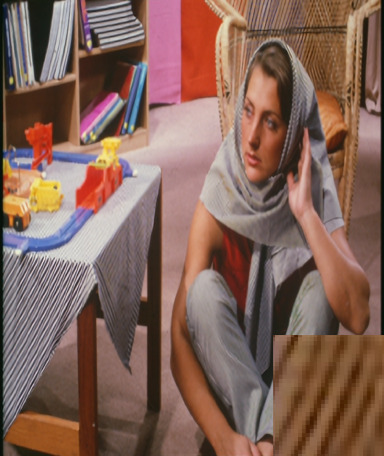}}
    \label{fig:barbara_DRRN}\hfill
    \subfloat[\scriptsize{MemNet}]{
        \includegraphics[scale=0.29]{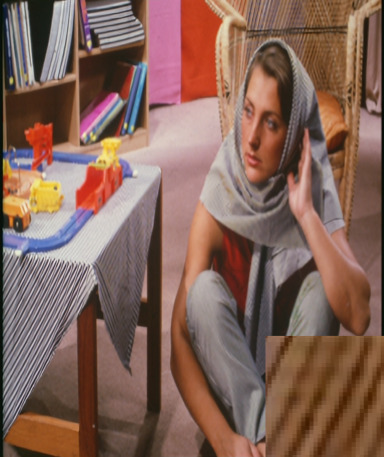}}
    \label{fig:barbara_memnet}\hfill
    \subfloat[\scriptsize{CARN}]{
        \includegraphics[scale=0.29]{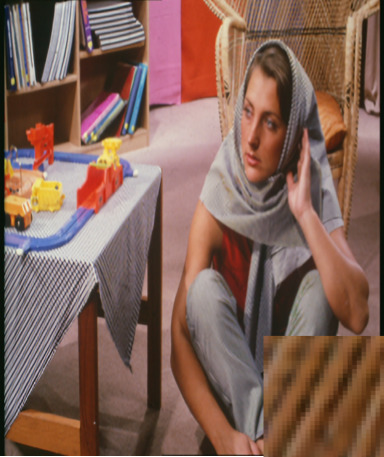}}
    \label{fig:barbara_carn}\hfill
    \subfloat[\scriptsize{BE-LCSC\_L}]{
        \includegraphics[scale=0.29]{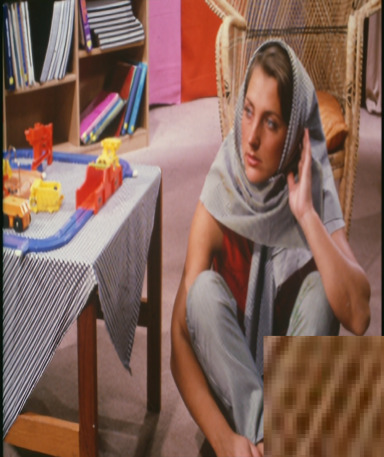}}
    \label{fig:barbara_LCSC}\hfill
    \caption{\small Results for upscaling factor ×3 on image Set14-barbara}
    \label{fig:barbara} 
\end{figure}
\begin{figure}[htbp]
    \centering
    \subfloat[\scriptsize{HR}]{
       \includegraphics[scale=0.29]{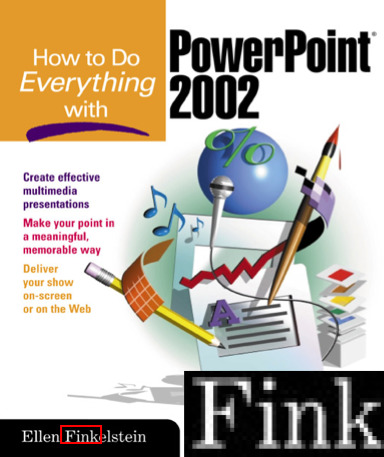}}
    \label{fig:ppt3_HR}\hfill
    \subfloat[\scriptsize{VDSR}]{
        \includegraphics[scale=0.29]{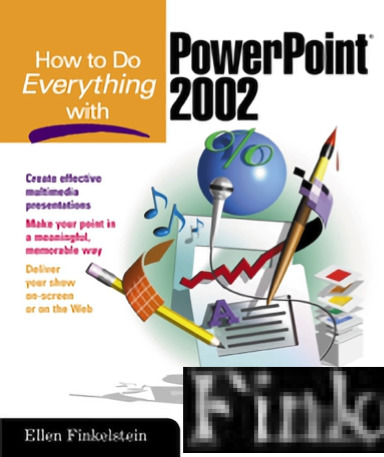}}
    \label{fig:ppt3_VDSR}\hfill
    \subfloat[\scriptsize{DRCN}]{
        \includegraphics[scale=0.29]{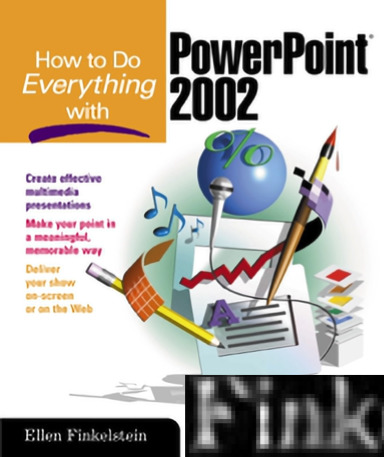}}
    \label{fig:ppt3_DRCN}\hfill
    \subfloat[\scriptsize{LapSR}]{
        \includegraphics[scale=0.29]{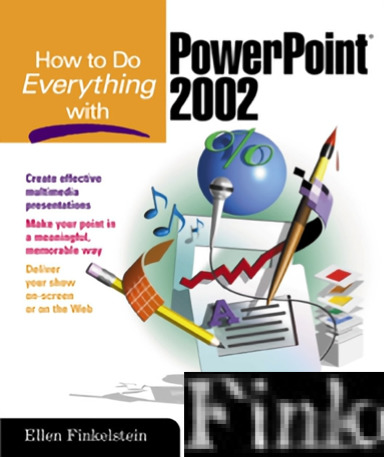}}
    \label{fig:ppt3_LapSR} \\
    \subfloat[\scriptsize{DRRN}]{
        \includegraphics[scale=0.29]{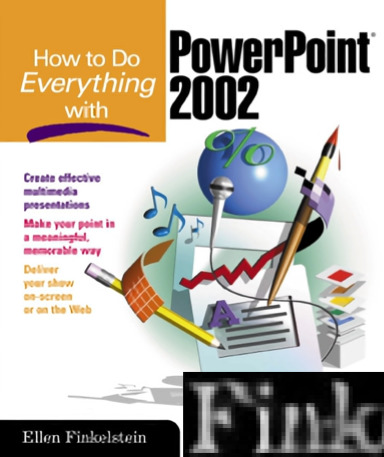}}
    \label{fig:ppt3_DRRN}\hfill
    \subfloat[\scriptsize{MemNet}]{
        \includegraphics[scale=0.29]{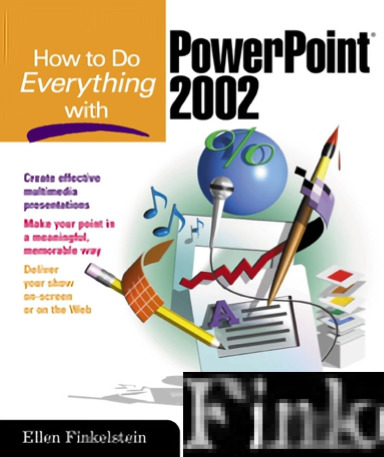}}
    \label{fig:ppt3_memnet}\hfill
    \subfloat[\scriptsize{CARN}]{
        \includegraphics[scale=0.29]{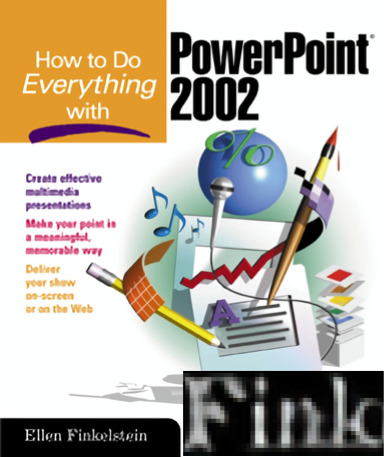}}
    \label{fig:ppt3_bic}\hfill
    \subfloat[\scriptsize{BE-LCSC\_L}]{
        \includegraphics[scale=0.29]{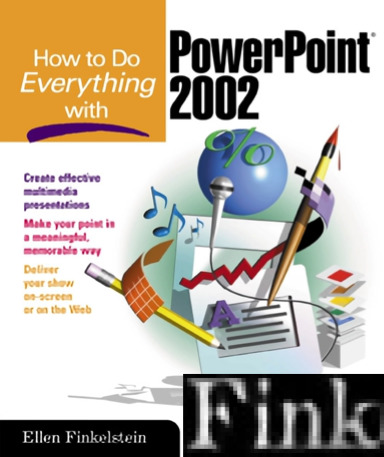}}
    \label{fig:ppt3_LCSC}\hfill
    \caption{\small Results for upscaling factor 3 on image Set14-ppt}
    \label{fig:ppt3} 
\end{figure}
\begin{figure}
    \centering
    \subfloat[\scriptsize{HR}]{
       \includegraphics[scale=0.181]{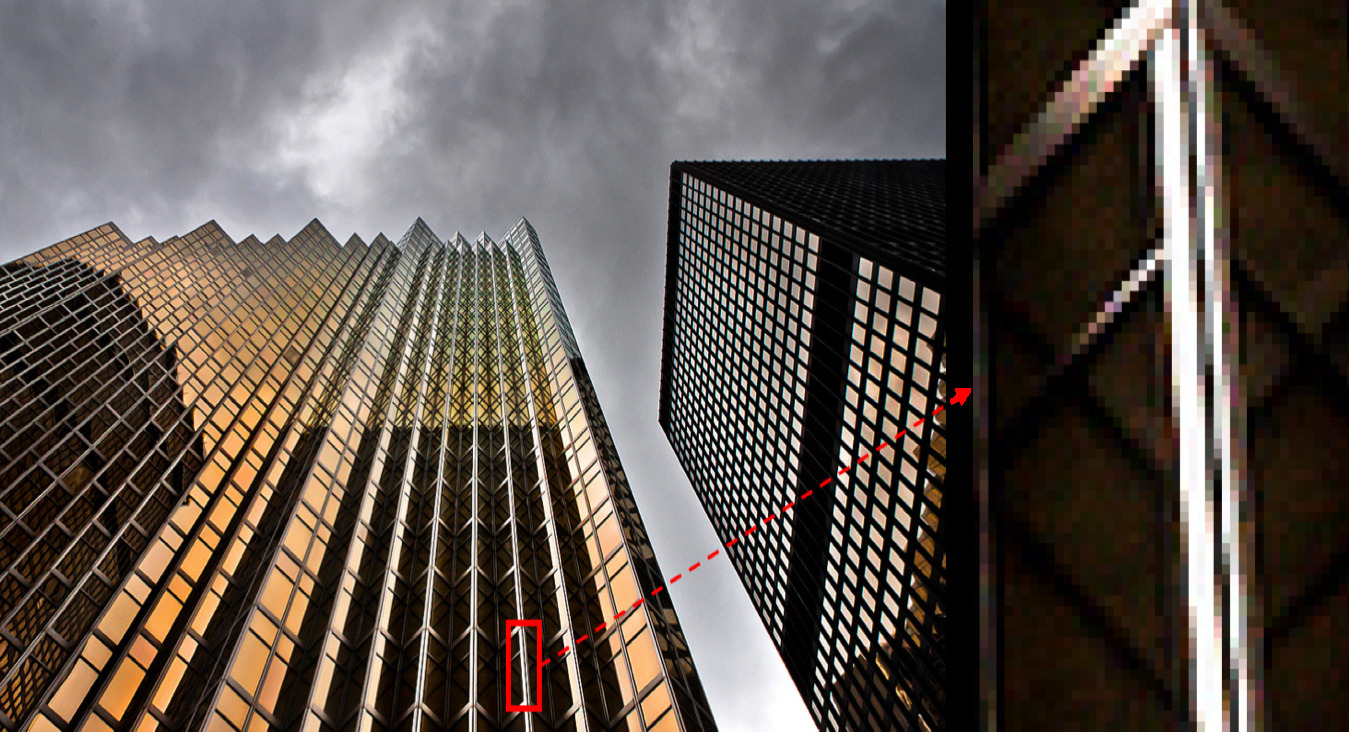}}
    \label{fig:img19_HR}\hfill
    \subfloat[\scriptsize{EDSR}]{
        \includegraphics[scale=0.181]{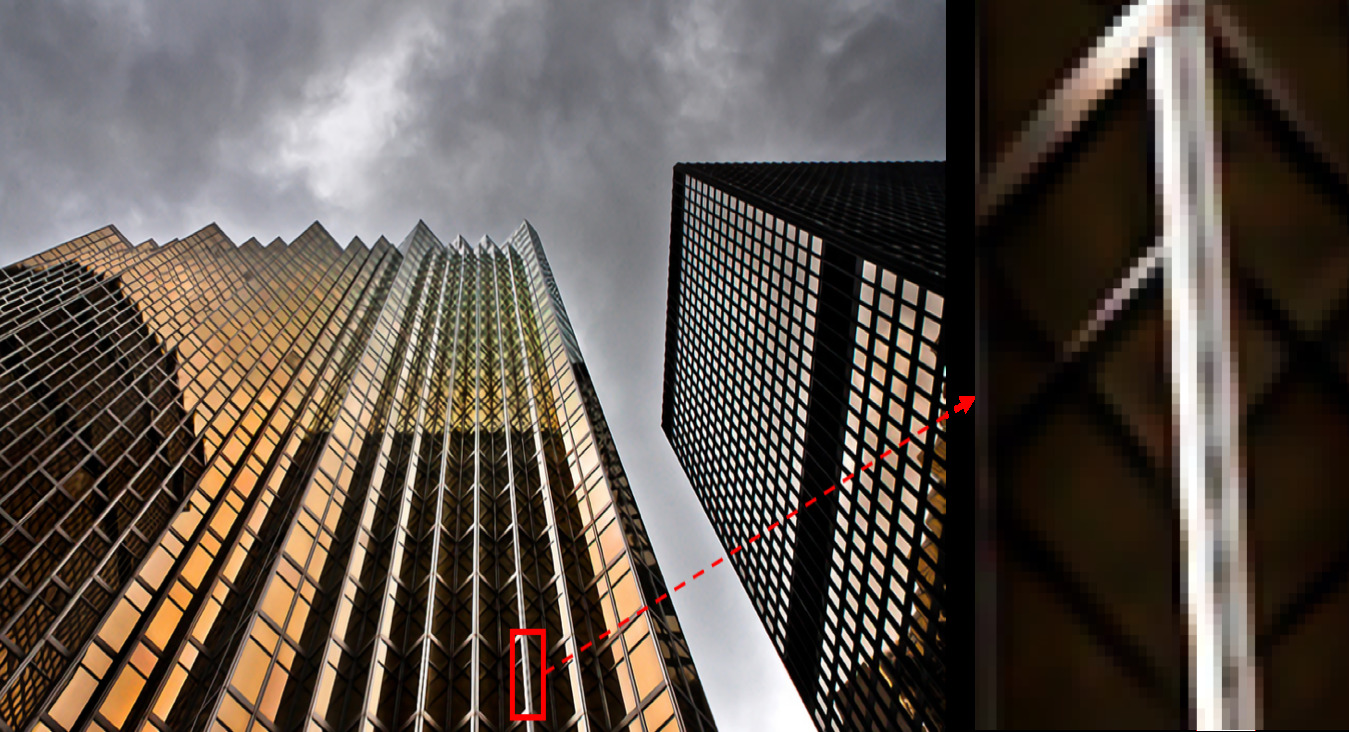}}
    \label{fig:img19_EDSR} \\
    \subfloat[\scriptsize{RDN}]{
        \includegraphics[scale=0.181]{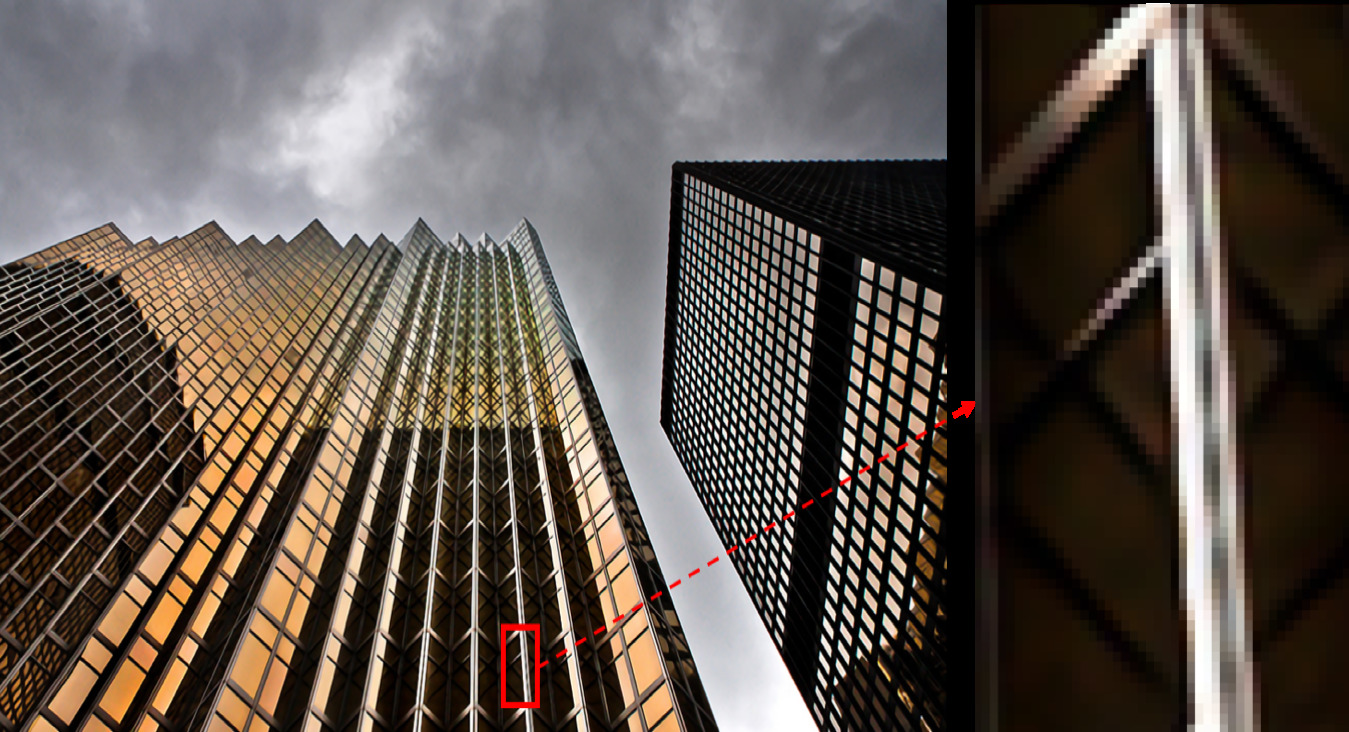}}
    \label{fig:img19_RDN}\hfill
    \subfloat[\scriptsize{E-LCSCNet}]{
        \includegraphics[scale=0.181]{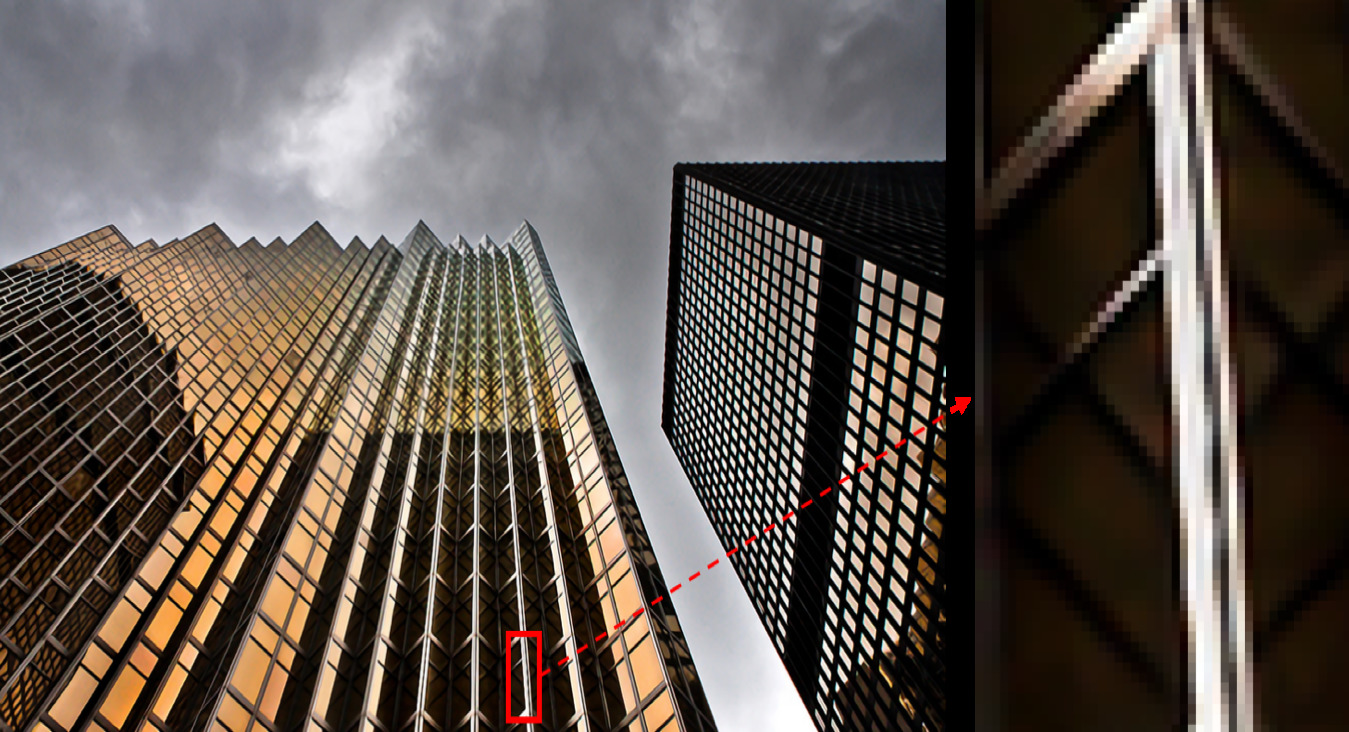}}
    \label{fig:img19_ELCSC}\\
    \caption{\small Results of large models for upscaling factor 3 on Urban100-img019}
    \label{fig:img19} 
\end{figure}

Quantitative comparisons on BE-LCSC\_L are listed in Table~\ref{chart:big}. Because operations in neural networks for SISR are mainly multiplication along with addition, we use the number of composite multiply-accumulate operations in CARN, denoted by Mult\&Adds, to measure computational efficiency, and we also assume that the HR image is $1280\times720$. From Table~\ref{chart:big}, we can see that among the models trained on 291, LCSC\_76\_291 achieves better accuracy than MemNet. As for efficiency, MemNet has fewer parameters due to its recursive structure, but LCSC\_76\_291 is more computation-efficient than MemNet. When compared with SelNet and CARN, BE-LCSC\_L is moderately computation-consuming but achieves obvious improvement. Among large models on DIV2K, our E-LCSCNet has the fewest Params for every scale. For $\times 2$ scale, our E-LCSCNet holds the same level with RDN but with a clear advantage in Mult\&Adds. For $\times 3$ and $\times 4$ scale, our E-LCSCNet performs better than EDSR and RDN, but is somehow more computation-consuming than RDN due to its fusion part. 

Representative qualitative comparisons are shown in Figs.\ref{fig:barbara}-\ref{fig:img19}. In Fig.\ref{fig:barbara}, our model restores the grid structure more precisely with fewer artifacts than other models. In Fig.\ref{fig:ppt3}, compared with blurry characters generated by other models, our result has sharper edges. In Fig.\ref{fig:img19}, compared with EDSR and RDN, E-LCSCNet recovers the line with the least blurry.   

\subsection{Implementation Details}
For training LCSCNet, we augment data ($90^{\circ}$, $180^{\circ}$ and $270^{\circ}$ rotation), and then downsample the LR input with the desired scaling factor. Like many methods trained on 291, we only take the luminance component for training. Ground truth for training is the residue between the bicubic of LR image and the original HR image, and all inputs are scaled into [-1, 1]. When trained on 291, training images are split into patches of sizes $18^2/36^2$, $12^2/36^2$ and $21^2/84^2$, respectively. We initialize all the convolution kernels as suggested by \cite{he2015delving}. All intermediate feature maps have 64 channels. For optimization, we use Adam~\cite{kingma2014adam} with its default settings. Learning rate is initialized as $10^{-4}$, and is divided by 10 every 15 epochs over the whole augmented dataset and the training is stopped after 60 epochs. For training, we use Keras~\cite{chollet2015keras}; for testing, we use MatConvNet~\cite{vedaldi2015matconvnet}. 

The training of BE-LCSC\_L and E-LCSCNet is based on the PyTorch~\cite{paszke2017pytorch} version of EDSR with the same setting of EDSR except that the batch size is 32, and training is terminated after 650 epochs. The codes are available from \url{https://github.com/XuechenZhang123/LCSC}.

\section{Conclusion and Future Works}\label{s:s_8}

In this paper, we propose the linear compressing based skip-connecting network (LCSCNet) for image SR, which combines the merits of the parameter-economic form of ResNet and the effective feature exploration of DenseNet. Linear compressing layers are adapted to implement skip connections, connecting former features and separating them from the newly-explored features. Compared with previous deep models with skip connections, our LCSCNet can explore relatively more new features with lower computational costs. Based on LCSCNet, to improve the performance of extremely deep and wide networks, the Enhanced LCSCNet is developed. An adaptive element-wise fusion strategy is also proposed, not only for further exploiting hierarchical information from diverse levels of deep models, but also for stabilizing the training deep models by adding extra paths for gradient flows. Comprehensive experiments and discussions are presented in this paper and demonstrate the rationality and superiority of the proposed methods. 

Future work can be mainly explored from the following two aspects:
1) it would be worthwhile to try to apply LCSCNet and E-LCSCNet or their basic units to other computer vision tasks;
and 2) in terms of Mult\&Adds in Table~\ref{chart:big}, we can see the computational cost for this part is still somewhat high despite that we have managed to control its complexity; therefore, further efforts can be made to further improve its efficiency.

\section*{Acknowledgment}
We would like to thank the authors of~\cite{kim2016accurate, kim2016deeply, LapSRN, tai2017image, Tai-MemNet-2017,ahn2018fast,lim2017enhanced,zhang2018residual} for releasing their source codes and models for comparison. We would also like to thank the Associate Editor and anonymous reviewers for their selfless dedication and constructive suggestions.

\bibliographystyle{IEEEtran}
\bibliography{LCSC}

\end{document}